\documentclass[letter,11pt]{article}
 \usepackage{jheppub}
 
 \usepackage{graphicx}
 \usepackage{hyperref}
 \usepackage[utf8]{inputenc}
 \usepackage{amssymb}
\usepackage{multirow}
\usepackage{soul}
 \usepackage{ulem}
 
 \usepackage{booktabs}
\usepackage{mathrsfs}
\usepackage{epsfig}
\usepackage{graphicx}
\usepackage{subfigure}
\usepackage{dcolumn}
\usepackage{bm}
\usepackage{amsmath}
\usepackage{slashed}
\usepackage{multirow}
\usepackage{color}
\usepackage{dcolumn}
\usepackage{bm}
\usepackage{color}
\usepackage[dvipsnames]{xcolor}

\def\beq{\begin{equation}}
\def\eeq{\end{equation}}
\def\eeqn{\end{equation}}
\newcommand\iden{\leavevmode\hbox{\small1\normalsize\kern-.33em1}}

\newcommand{\bea} {\begin{eqnarray}}
\newcommand{\eea} {\end{eqnarray}}

\def\tb {t_\beta}
\def\sb  {s_{\beta}}
\def\cb  {c_{\beta}}
\def\stwob  {s_{2\beta}}

\def\bra{\langle}
\def\ket{\rangle}

\let\jnfont=\rm
\def\NPB#1 {{\jnfont Nucl.\ Phys.\ B }{\bf #1} }
\def\PLB#1 {{\jnfont Phys.\ Lett.\ B }{\bf #1} }
\def\EPJC#1 {{\jnfont Eur.\ Phys.\ Jour.\ C }{\bf #1} }
\def\PRD#1 {{\jnfont Phys.\ Rev.\ D }{\bf #1} }
\def\PRL#1 {{\jnfont Phys.\ Rev.\ Lett.\ }{\bf #1} }
\def\MPLA#1 {{\jnfont Mod.\ Phys.\ Lett.\ A }{\bf #1} }
\def\JPG#1 {{\jnfont J.\ Phys.\ G }{\bf #1} }
\def\CTP#1 {{\jnfont Commun.\ Theor.\ Phys.\ }{\bf #1} }
\def\JHEP#1 {{\jnfont JHEP \ }{\bf #1} }
\def\NPPS#1 {{\jnfont Nucl.\ Phys.\ Proc.\ Suppl.\ }{\bf #1} }
\def\CPC#1 {{\jnfont Comput.\ Phys.\ Commun.\ }{\bf #1} }
\def\CPL#1 {{\jnfont Chin.\ Phys.\ Lett. }{\bf #1} }
\def\APPB#1 {{\jnfont Acta\ Phys.\ Polon.\ B }{\bf #1} }

\def\lsim{\raise0.3ex\hbox{$<$\kern-0.75em\raise-1.1ex\hbox{$\sim$}}}
\def\gsim{\raise0.3ex\hbox{$>$\kern-0.75em\raise-1.1ex\hbox{$\sim$}}}
\def\PR#1 {{\jnfont Phys.\ Rept. }{\bf #1} }
\def\CHC#1 {{\jnfont Chin.\ Phys.\ C }{\bf #1} }
\def\NIMA#1 {{\jnfont Nucl.\ Instrum.\ Meth.\ A }{\bf #1} }
\def\JCAP#1 {{\jnfont JCAP \ }{\bf #1} }
\def\ASA#1 {{\jnfont Astron.\ Astrophys.\ A }{\bf #1} } 
\def\chinpc#1,{{\jnfont Chin.\ Phys.\ C }{\bf #1},} 
\def\half{\tfrac{1}{2}}

\title{Two-Higgs-doublet models in light of current experiments:  a brief review}

\author{Lei Wang$^{1}$, Jin Min Yang$^{2,3}$, Yang Zhang$^{4}$}
\affiliation{$^1$ Department of Physics, Yantai University, Yantai 264005, P. R. China\\
$^2$ CAS Key Laboratory of Theoretical Physics,
     Institute of Theoretical Physics,  Chinese Academy of Sciences, Beijing 100190, P. R. China\\
$^3$ School of Physical Science, University of Chinese Academy of Sciences, Beijing 100049, P. R. China\\
$^4$ School of Physics and Microelectronics, Zhengzhou University, ZhengZhou 450001, P. R. China}

\abstract{
We briefly survey several typical CP-conserving two-Higgs-doublet models (2HDMs) 
in light of current experiments. 
First we derive the masses 
and couplings of the mass eigenstates from the Lagrangians. Then we analyze the constraints from theory and 
oblique electroweak parameters. Finally, we delineate the status of 2HDM in light of the LHC searches, 
the dark matter detections and the muon $g-2$ measurement.} 

\begin{document}
\maketitle 
\indent
\newpage

\section{Introduction}
A two-Higgs-doublet model (2HDM) is a simple extension of the Standard Model (SM) by introducing an additional $SU(2)_L$
Higgs doublet, which predicts three neutral Higgs bosons and a pair of 
charged Higgs bosons $H^{\pm}$. The 2HDM can be CP-violating \cite{t.lee}, but it is also useful to study its CP conserving version, 
where the neutral Higgs bosons can be classified into the CP-even states ($h$ and $H$) and the CP-odd state ($A$).
 The tree-level flavour changing neutral current (FCNC) can appear in the general 2HDM, which is forbidden by imposing $Z_2$ discrete symmetry in several different ways,
 such as type-I 2HDM \cite{i-1,i-2}, 
type-II 2HDM \cite{i-1,ii-2}, lepton-specific 2HDM (L2HDM), flipped 2HDM \cite{xy-1,xy-2,xy-3,xy-4,xy-5,xy-6}, 
and inert 2HDM \cite{inert2h,inert2h-2,inert2h-3}.
Also the tree-level FCNC is absent in the aligned 2HDM in which the Yukawa-coupling matrices of the two Higgs doublet fields are assumed to be proportional \cite{aligned2h}. 
In addition, due to certain type of symmetry, the FCNC is naturally suppressed by the off-diagonal element of the CKM matrix in the Branco-Grimus-Lavoura (BGL) 2HDM \cite{bgl2hdm}.

Various 2HDMs have been extensively studied in particle physics.
Because of the plenty Yukawa couplings of the quarks and leptons, the 2HDMs have been 
studied in meson decays, and the L2HDM is used to explain the muon $g-2$ anomaly \cite{mu2h5,mu2h6,mu2h8,mu2h9,mu2h10,mu2h11,
mu2h15,mu2h16,mu2h16-2,mu2h16-3,mu2h16-4,mu2h17,mu2h18,mu2h19,mu2h19-2,mu2h20,mu2h21,mu2h22}. 
In the inert 2HDM, one may take the lightest component of the inert Higgs doublet field is neutral, and consider it as a dark matter (DM) candidate because of its stability. 
If an additional field protected by a new symmetry is added to other types of 2HDMs, then these models also can provide a DM candidate.
In these models, the multiple scalar fields can be as the portals between the DM and the SM sector, and
lead to some interesting effects on the DM observables via their various Yukawa couplings \cite{2hisos-0,2hisos-1,2hisos-2,2hisos-3,2hisos-4,2hisos-5,2hisos-6,dmbu,1708.06882,1608.00421,1801.08317,1808.02667,2102.01588,blind-dm-1,blind-dm-2,l2hdm-dm-1,2112.15536}.
On the other hand, the analyses of ATLAS and CMS collaborations at the LHC show that the properties of the discovered 125 GeV Higgs boson 
agree well with the SM Higgs boson \cite{cmsh,atlh}. 
Other than that, no experiment claims to have observed any new resonance with $5\sigma$ level. However, there are some interesting excesses which
imply the existence of new scalars. For example, the CMS Run II results for Higgs boson searches in the diphoton final state show a local excess of $\sim$ $3\sigma$ around 96 GeV
 \cite{cms96}. The ATLAS collaboration reported a local excess of $\sim$ $3\sigma$ around 130 GeV in the searches for $t\to H^\pm b$ with $H^\pm\to cb$ \cite{atlas130}.
Besides, very recently the CDF II result for the $W$-mass has an approximate $7\sigma$ discrepancy from the SM prediction [55]. The 2HDM can 
give additional corrections to the masses of gauge bosons via the self-energy diagrams exchanging extra Higgs fields, and simply explain
the CDF $W$-mass when the extra Higgses have appropriate mass splittings (see e.g., \cite{mw2h-1,mw2h-2,mw2h-3,mw2h-4,mw2h-5,mw2h-6,mw2h-7,mw2h-8,mw2h-9,mw2h-10,mw2h-11,mw2h-12,mw2h-13,mw2h-14}).

In the literature there already have been some reviews on 2HDMs (see e.g., \cite{1106.0034,1507.06424,1609.06089,2203.01396,review-4,review-5}). 
In this note we emphasize current experiments and briefly review several typical CP-conserving 2HDMs.
We will start from the Lagrangians and derive the masses and couplings 
of the particles. Then we analyze the constraints from theory and oblique electroweak parameters, respectively. 
Finally, we discuss the status of 2HDMs in light of the LHC searches, the dark matter detections 
and the muon $g-2$ measurement. 

The content is organized as follows. In Sections II and III, we demonstrate several typical CP-conserving 2HDMs and discuss 
the constraints from theory and oblique parameters. In Sections IV, V, and VI,  we review the status of
the 2HDMs in light of the LHC Higgs searches, the DM detections and the muon $g-2$ measurement.
Finally, we give a summary in Section \ref{summary}. 

\section{Several typical 2HDMs}
The general scalar potential of 2HDM is given as
\begin{eqnarray} \label{V2HDM} V_{\rm tree} &=& m_{11}^2
(\Phi_1^{\dagger} \Phi_1) + m_{22}^2 (\Phi_2^{\dagger}
\Phi_2) - m_{12}^2 (\Phi_1^{\dagger} \Phi_2 + \rm h.c.)\nonumber \\
&&+ \frac{\lambda_1}{2}  (\Phi_1^{\dagger} \Phi_1)^2 +
\frac{\lambda_2}{2} (\Phi_2^{\dagger} \Phi_2)^2 + \lambda_3
(\Phi_1^{\dagger} \Phi_1)(\Phi_2^{\dagger} \Phi_2) + \lambda_4
(\Phi_1^{\dagger}
\Phi_2)(\Phi_2^{\dagger} \Phi_1) \nonumber \\
&&+ \left[\frac{\lambda_5}{2} (\Phi_1^{\dagger} \Phi_2)^2 + \lambda_6 (\Phi_1^{\dagger} \Phi_1)(\Phi_1^{\dagger} \Phi_2)
+\lambda_7 (\Phi_2^{\dagger} \Phi_2)(\Phi_1^{\dagger} \Phi_2) +\rm
h.c.\right],
\end{eqnarray}
 and the $\Phi_1$ and $\Phi_2$ are complex Higgs doublets with hypercharge $Y = 1$:
\begin{equation}
\Phi_1=\left(\begin{array}{c} \phi_1^+ \\
\frac{1}{\sqrt{2}}\,(v_1+\phi_1+\textrm{i}a_1)
\end{array}\right)\,, \ \ \
\Phi_2=\left(\begin{array}{c} \phi_2^+ \\
\frac{1}{\sqrt{2}}\,(v_2+\phi_2+\textrm{i}a_2)
\end{array}\right).
\end{equation}
Here we restrict to the CP-conserving models in which all $\lambda_i$ and
$m_{12}^2$ are real and the electroweak vacuum expectation values (VEVs) $v_1$ and $v_2$ are also real with $v^2 = v^2_1 + v^2_2 = (246~\rm GeV)^2$.

\subsection{Type-I, type-II, lepton-specific and flipped 2HDMs}
In order to forbid tree-level FCNC, one may introduce an additional $Z_2$ discrete
symmetry under which the charge assignments of fields are shown in Table~\ref{tab:3_models}.
Because of this $Z_2$ symmetry, the $\lambda_6$ and $\lambda_7$ terms in the general scalar potential in Eq. (\ref{V2HDM})
are absent, while the soft breaking $m_{12}^2$ term is still allowed. The mass parameters $m^{2}_{11}$ and $m^{2}_{22}$ 
in the potential are determined by the potential minimization conditions at $(v_1,v_2)$:
\beq
\begin{split}
&\quad m_{11}^2 = m_{12}^2 \tb - \frac{1}{2} v^2 \left( \lambda_1 \cb^2 + \lambda_{345}\sb^2 \right)\,,\\
& \quad m_{22}^2 =  m_{12}^2 / \tb - \frac{1}{2} v^2 \left( \lambda_2 \sb^2 + \lambda_{345}\cb^2 \right)\,,
\end{split}
\label{min_cond}
\eeq
where the shorthand notations  $\tb\equiv \tan\beta=v_2/v_1$, $\sb\equiv \sin\beta$, $\cb \equiv \cos\beta$,
and $\lambda_{345} = \lambda_3+\lambda_4+\lambda_5$ are employed.

\begin{table}
\begin{center}
\caption{The $Z_2$ charge assignment in the four types of 2HDMs without FCNC. The other fields are even under $Z_2$ symmetry.}
\label{tab:3_models}
\begin{tabular}{|c|c|c|c|c|c|} \hline
Model & $~~\Phi_2~~$ & ~~$\Phi_1$ ~~&~~$u_R^i$~~     &~~ $d_R^i$~~& ~~$e_R^i$ ~~    \\
\hline
Type I     & $+$   &$ -$ & $+$  & $+$ & $+$    \\
\hline
Type II    & $+$   &$-$  & $+$  & $-$ & $-$   \\
\hline
Lepton-specific    &$+$  & $-$  & $+$ & $+$   & $-$    \\
\hline
Flipped    & $+$   &$-$  & $+$  & $-$ & $+$    \\
\hline
\end{tabular}
\end{center}
\end{table}

From the scalar potential in Eq. (\ref{V2HDM}) with $\lambda_6=\lambda_7=0$, we can obtain
the mass matrices of the Higgs fields 
\bea
&&\label{Z2_basis_diag}
\begin{pmatrix}\phi_1 & \phi_2\end{pmatrix} \begin{pmatrix} m_{12}^2 \tb+ \lambda_1 v^2 \cb^2  &\quad - m^2_{12} + {\lambda_{345} \over 2} v^2 \stwob \\[5pt]
- m^2_{12} + {\lambda_{345} \over 2} v^2 \stwob & \quad m_{12}^2/ \tb+ \lambda_2 v^2 \sb^2  \\[5pt] \end{pmatrix} \begin{pmatrix}\phi_1\\ \phi_2\end{pmatrix}\,,\label{eq:mass_matrix}\\
&&\begin{pmatrix}a_1 & a_2\end{pmatrix} \left[ m_{12}^2 - {1\over 2} \lambda_5 v^2 \stwob \right] \begin{pmatrix} \tb & \quad -1 \\[5pt]
-1 & \quad  1/\tb \end{pmatrix} \begin{pmatrix}a_1\\ a_2\end{pmatrix}\,,\\
&&\begin{pmatrix}\phi^+_1 & \phi^+_2\end{pmatrix} \left[ m_{12}^2 - {1\over 4} (\lambda_4+ \lambda_5) v^2 \stwob \right] \begin{pmatrix} \tb & \quad -1 \\[5pt]
-1 & \quad  1/\tb \end{pmatrix} \begin{pmatrix}\phi^-_1 \\ \phi^-_2\end{pmatrix}\,.
\eea
The mass eigenstates are obtained from the original fields by the rotation matrices:
\begin{eqnarray}
\left(\begin{array}{c}H \\ h \end{array}\right) =  \left(\begin{array}{cc}\cos\alpha & \sin\alpha \\ -\sin\alpha & \cos\alpha \end{array}\right)  \left(\begin{array}{c} \phi_1 \\ \phi_2 \end{array}\right) , \\
\left(\begin{array}{c}G^0 \\ A \end{array}\right) =  \left(\begin{array}{cc}\cos\beta & \sin\beta \\ -\sin\beta & \cos\beta \end{array}\right)  \left(\begin{array}{c} a_1 \\ a_2 \end{array}\right) , \\
\left(\begin{array}{c}G^{\pm} \\ H^{\pm} \end{array}\right) =  \left(\begin{array}{cc}\cos\beta & \sin\beta \\ -\sin\beta & \cos\beta \end{array}\right)  \left(\begin{array}{c} \phi^{\pm}_1 \\ \phi^{\pm}_2 \end{array}\right),
\end{eqnarray}
where $G^0$ and $G^\pm$ are Goldstone bosons which are absorbed as longitudinal components of the $Z$ and $W^\pm$ bosons. 
The remained physical states are two neutral
CP-even states $h$ and $H$, one neutral pseudoscalar $A$, and a pair of charged scalars $H^{\pm}$. 
Their masses are given by 
\begin{align}
m^2_{H,h} &= \frac{1}{2}\left[M^2_{P, 11} + M^2_{P,22}\pm \sqrt{(M^2_{P,11}-M^2_{P,22})^2+4 (M^2_{P,12})^2 } \right]  \label{eq:hmass} \ ,\\
m_A^2 &= \frac{m_{12}^2}{s_\beta c_\beta} - \lambda_5 v^2 \label{eq:Amass} \ , \\
m_{H^{\pm}}^2 &= \frac{m_{12}^2}{s_\beta c_\beta} - \frac{1}{2} (\lambda_4+\lambda_5) v^2 \ ,\label{eq:Hcmass}
\end{align}
where $M^2_{P}$ is the mass matrix shown in Eq.~\ref{eq:mass_matrix}.

The gauge-kinetic Lagrangian is given as
\beq
\mathcal{L}_\mathrm{g} = \left( D^\mu \Phi_1 \right)^\dagger \left( D_\mu \Phi_1 \right) + \left( D^\mu \Phi_2 \right)^\dagger \left( D_\mu \Phi_2 \right).
\eeq
We can obtain the neutral Higgs couplings to $VV$ ($VV\equiv ZZ, WW$)
\begin{align}
\mathcal{L}_\mathrm{g} \supset &\frac{g^2+g'^2}{8}v^2~ZZ~\left(1+2\frac{h}{v}y_h^V+2\frac{H}{v}y_H^V\right) \nonumber\\
                               &+\frac{g^2}{4}v^2~W^+W^-~\left(1+2\frac{h}{v}y_h^V+2\frac{H}{v}y_H^V\right),
\end{align}
where $y_h^V=\sin(\beta-\alpha)$ and $y_H^V=\cos(\beta-\alpha)$.
 
According to different charge assignments, there are four different models with Yukawa interactions:  
 \bea
&&- {\cal L} = Y_{u2}\,\overline{Q}_L \, \tilde{{ \Phi}}_2 \,u_R
+\,Y_{d2}\,
\overline{Q}_L\,{\Phi}_2 \, d_R\, + \, Y_{\ell 2}\,\overline{L}_L \, {\Phi}_2\,e_R+\, \mbox{h.c.}\, ({\rm  ~type~ I}),\\
&&
- {\cal L} = Y_{u2}\,\overline{Q}_L \, \tilde{{ \Phi}}_2 \,u_R
+\,Y_{d1}\,
\overline{Q}_L\,{\Phi}_1 \, d_R\, + \, Y_{\ell 1}\,\overline{L}_L \, {\Phi}_1\,e_R+\, \mbox{h.c.}\, ({\rm ~type~ II}),\\ 
&&
- {\cal L} = Y_{u2}\,\overline{Q}_L \, \tilde{{ \Phi}}_2 \,u_R
+\,Y_{d1}\,
\overline{Q}_L\,{\Phi}_2 \, d_R\, + \, Y_{\ell 1}\,\overline{L}_L \, {\Phi}_1\,e_R+\, \mbox{h.c.}\, ({\rm ~lepton~ specific}),\\
&&
- {\cal L} = Y_{u2}\,\overline{Q}_L \, \tilde{{ \Phi}}_2 \,u_R
+\,Y_{d1}\,
\overline{Q}_L\,{\Phi}_1 \, d_R\, + \, Y_{\ell 1}\,\overline{L}_L \, {\Phi}_2\,e_R+\, \mbox{h.c.}\, ({\rm ~flipped~ model}),
\eea 
 where
$Q_L^T=(u_L\,,d_L)$, $L_L^T=(\nu_L\,,l_L)$, $\widetilde\Phi_{1,2}=i\tau_2 \Phi_{1,2}^*$, and $Y_{u2}$, 
$Y_{d1,2}$ and $Y_{\ell 1,2}$ are $3 \times 3$ matrices in family space.

We can obtain the Yukawa couplings 
\begin{eqnarray}
- {\cal L}_Y &=& \frac{m_f}{v}~ y_h^f~h\bar{f}f+\frac{m_f}{v}~y_H^f~H\bar{f}f\nonumber\\
&&-i\frac{m_u}{v}\kappa_u ~A \bar{u} \gamma_5 u + i\frac{m_d}{v}\kappa_d~A \bar{d} \gamma_5 d+ i\frac{m_\ell}{v}\kappa_\ell~ A \bar{\ell} \gamma_5 \ell\nonumber\\
&&+ H^+~ \bar{u} ~V_{\rm CKM}~ (\frac{\sqrt{2}m_d}{v}\kappa_d P_R  - \frac{\sqrt{2}m_u}{v}\kappa_u P_L)  d+ h.c.\nonumber\\
& &+ \frac{\sqrt{2}m_\ell}{v}\kappa_\ell H^+~\bar{\nu} P_R  e + h.c.\label{yuka-coupling}
\end{eqnarray}
where $y_h^f=\sin(\beta-\alpha)+\cos(\beta-\alpha)\kappa_f$ and $y_H^f=\cos(\beta-\alpha)-\sin(\beta-\alpha)\kappa_f$.
The values of $\kappa_u$, $\kappa_d$ and $\kappa_\ell$ for the four models are shown in Table ~\ref{tab:types}.

\begin{table}
\centering
 \caption{The $\kappa_u$, $\kappa_d$, and $\kappa_\ell$ for the four types of 2HDMs.} 
\vspace{0.2cm} 
\label{tab:types}
\begin{tabular}{cccccc}
  \hline               
          &~~~ type-I~~~ & ~~~type-II~~~ & lepton-specific~ &~ flipped \\
         \hline
         $\kappa_u$ & $1/\tb$  & $1/\tb$ & $1/\tb$ & $1/\tb $\\
         $\kappa_d$ & $1/\tb$  & $-\tb$ & $1/\tb$ & $-\tb $\\
         $\kappa_\ell$ & $1/\tb$  & $-\tb$  & $-\tb$ & $1/\tb $\\
         \hline
\end{tabular}
\end{table}

\subsection{Inert Higgs doublet model}
We impose an exact $Z_2$ discrete symmetry in the 2HDM and assume that it remains after the potential minimization. Under the $Z_2$ symmetry all the SM fields are taken to be even, 
while the new (inert) doublet $\Phi_2$ is odd:
\begin{equation}
\Phi_1=\left(\begin{array}{c} G^+ \\
\frac{1}{\sqrt{2}}\,(v+h+\textrm{i}G)
\end{array}\right)\,, \ \ \
\Phi_2=\left(\begin{array}{c} H^+ \\
\frac{1}{\sqrt{2}}\,(H+\textrm{i}A)
\end{array}\right).
\end{equation}
The $\Phi_1$ field has a vev $v=246$ GeV, and $\Phi_2$ has no vev.

 The scalar potential is  
\begin{eqnarray}  \mathcal{V} &=& m_{11}^2
(\Phi_1^{\dagger} \Phi_1) + m_{22}^2 (\Phi_2^{\dagger}
\Phi_2) + \frac{\lambda_1}{2}  (\Phi_1^{\dagger} \Phi_1)^2 +
\frac{\lambda_2}{2} (\Phi_2^{\dagger} \Phi_2)^2 \nonumber \\
&& + \lambda_3
(\Phi_1^{\dagger} \Phi_1)(\Phi_2^{\dagger} \Phi_2) + \lambda_4
(\Phi_1^{\dagger}
\Phi_2)(\Phi_2^{\dagger} \Phi_1) + \left[\frac{\lambda_5}{2} (\Phi_1^{\dagger} \Phi_2)^2 + \rm
h.c.\right].
\end{eqnarray}
The parameter $m_{11}^2$ is fixed by the scalar potential minimum conditions
\beq \label{sminconds}
m_{11}^2=-\half \lambda_1 v^2.
\eeq
The fields $H^\pm$ and $A$ are the mass eigenstates and their masses are given by
\beq \label{masshp}
 m_{H^\pm}^2  = m_{22}^2+\frac{\lambda_3}{2} v^2, ~~m_{A}^2  = m_{H^\pm}^2+\frac{1}{2}(\lambda_4-\lambda_5) v^2.
 \eeq
There is no mixing between $h$ and $H$, which are the CP-even mass eigenstates 
\beq \label{massh}
 m_{h}^2  = \lambda_1 v^2\equiv (125~{\rm GeV })^2, ~~m_{H}^2  = m_{A}^2+\lambda_5 v^2.
 \eeq

The fermion masses can be obtained via the Yukawa interactions with $\Phi_1$ 
\beq \label{yukawacoupling} 
- {\cal L} = y_u\,\overline{Q}_L \,
\tilde{{ \Phi}}_1 \,u_R +\,y_d\,
\overline{Q}_L\, {\Phi}_1 \, d_R  \, + \, y_l\,\overline{L}_L \, {\Phi}_1
\,e_R \,+\, \mbox{h.c.}\,, 
\eeq
where $y_u$, $y_d$ and $y_\ell$ are $3 \times 3$ matrices in family space. 
Because of the exact $Z_2$ symmetry, the inert field $\Phi_2$ has no Yukawa interactions 
with fermions. 
The lightest neutral field,  $H$ or $A$, is stable and may be considered as a DM candidate.
If right-handed neutrinos are introduced, then $\Phi_2$ can interact with them, giving rise to the neutrino masses via the one loop with DM \cite{mamodel}.

\section{Constraints from theory and oblique parameters}
\subsection{Vacuum stability}
Vacuum stability requires the potential to be bounded from below 
and stay positive for arbitrarily large values of the fields.
The Higgs potential with a soft $Z_2$ symmetry breaking term is given by
\begin{eqnarray} \label{V2HDMii} 
V_{\rm tree} &=& m_{11}^2
(\Phi_1^{\dagger} \Phi_1) + m_{22}^2 (\Phi_2^{\dagger}
\Phi_2) - \left[m_{12}^2 (\Phi_1^{\dagger} \Phi_2 + \rm h.c.)\right]\nonumber \\
&&+ \frac{\lambda_1}{2}  (\Phi_1^{\dagger} \Phi_1)^2 +
\frac{\lambda_2}{2} (\Phi_2^{\dagger} \Phi_2)^2 + \lambda_3
(\Phi_1^{\dagger} \Phi_1)(\Phi_2^{\dagger} \Phi_2) + \lambda_4
(\Phi_1^{\dagger}
\Phi_2)(\Phi_2^{\dagger} \Phi_1) \nonumber \\
&&+ \left[\frac{\lambda_5}{2} (\Phi_1^{\dagger} \Phi_2)^2 + \rm
h.c.\right].
\end{eqnarray}
The fields can be parametrized as 
\beq
\Phi_1^\dagger \Phi_1=X_1^2,~~~\Phi_2^\dagger \Phi_2=X_2^2,~~~\Phi_1^\dagger\Phi_2=X_1X_2\rho {\rm e}^{i\theta} ~{\rm with}~0\leq\rho\leq 1. 
\eeq
For large values of the fields, the quadratic terms can be neglected and the quartic part is 
\beq  V_{4} = \frac{\lambda_1}{2}  X_1^4 +
\frac{\lambda_2}{2} X_2^4 + \lambda_3 X_1^2X_2^2 + \lambda_4  X_1^2X_2^2\rho^2 + \lambda_5 X_1^2X_2^2\rho^2\cos 2\theta.
\eeq
After stabilizing $\theta$ at the minimum, we obtain the $\theta$-independent
part of potential
 \beq  V_{\theta-{\rm indep}} = \frac{\lambda_1}{2}  X_1^4 +
\frac{\lambda_2}{2} X_2^4 + \lambda_3 X_1^2X_2^2 + \lambda_4  X_1^2X_2^2\rho^2 - \mid\lambda_5\mid X_1^2X_2^2\rho^2.
\eeq
For $\lambda_4-\mid\lambda_5\mid >$ 0, the potential has a minimal value at $\rho=0$,
 \beq  V_{\theta-\rho-{\rm indep}} = \frac{\lambda_1}{2}  X_1^4 +
\frac{\lambda_2}{2} X_2^4 + \lambda_3 X_1^2X_2^2.\eeq
For $\lambda_4-\mid\lambda_5\mid <$ 0, the potential has a minimal value at $\rho=1$,
 \beq  V_{\theta-\rho-{\rm indep}} = \frac{\lambda_1}{2}  X_1^4 +
\frac{\lambda_2}{2} X_2^4 + \lambda_3 X_1^2X_2^2 + \lambda_4  X_1^2X_2^2 - \mid\lambda_5\mid X_1^2X_2^2.
\eeq
Therefore, the vacuum stability requires
\beq
\lambda_1 > 0,~~\lambda_2 > 0,~~\lambda_3 + \sqrt{\lambda_1\lambda_2} > 0,~~\lambda_3 + \lambda_4 - \mid\lambda_5\mid +\sqrt{\lambda_1\lambda_2} > 0.
\eeq

In addition, there is the possibility that the 2HDM scalar potential of Eq. (\ref{V2HDMii}) has two minima, and the selected minimum is required to be global in order to
 avoid a metastable vacuum, which imposes the following condition \cite{1303.5098},
\beq
m_{12}^2 (m_{11}^2-k^2 m_{22}^2)(\tan\beta-k) > 0
\eeq
with $k=\sqrt[4]{\lambda_1/\lambda_2}.$

\subsection{Unitarity}
The amplitudes for scalar-scalar scattering $s_1 s_2 \to s_3 s_4$
at high energies respect unitarity \cite{unit-1}. A simple and explicit derivation can also be 
found in \cite{unit-2}. 
The starting point is the unitarity of the $S$ matrix, $S=1+iT$,
\begin{align}
S S^\dagger = 1 \longrightarrow T^\dagger  T = - i(T-T^\dagger).
\label{EQ:simpleSS}\end{align}
Then in terms of matrix elements of scattering from a pair of particles $a = {1, 2}$ with momenta ${p_1, p_2}$ to a pair $b = {3, 4}$ with
momenta ${k_3, k_4}$ we have 
\begin{align}
\bra \{k,b \} |iT| \{p,a\}\ket \equiv  i \mathcal{M}_{ba}(2\pi)^4 \delta^4 (k_3+k_4-p_1-p_2).
\end{align}
We can obtain a bound on the partial wave
\begin{align}\label{apartp}
  - \frac{i}{2} (a_J - a_J^{\dagger}) \ge   a_J a_J^{\dagger},
\end{align}
where $a_J$ is a normal matrix related to the partial wave decomposition of $2 \rightarrow 2$ scattering matrix elements $\mathcal{M}_{ba}$,
\begin{align}
a_J^{\rm ba} \equiv& \frac{1}{32\pi} \sqrt{\frac{4 |\mathbf{p}^b| |\mathbf{p}^a|}{2^{\delta_{12}} 2^{\delta_{34}}\, s}} \int_{-1}^1 d(\cos \theta) \mathcal{M}_{ba} (\cos \theta) P_J (\cos \theta).
\end{align}
The factor  $\delta_{12} (\delta_{34})$ is $1$ when the particles $1$ and $2$ ($3$ and $4$) are identical,
 and zero otherwise. $P_J$ are the Legendre polynomials, $\mathbf{p}^i$ is the 
centre-of-mass three-momentum for particle $i$, and $s=(p_1 + p_2)^2$ is the standard Mandelstam variable.

We can diagonalize $a$ and $a^\dagger$ in Eq. (\ref{apartp}) with an unitary matrix, 
and obtain the constraints on  the eigenvalues $(a_J^i)$:
\beq
\mathrm{Im} (a_J^i) \ge | a_J^i|^2 \to
  \left[(\mathrm{Re} (a_J^i) \right]^2 + \left[ (\mathrm{Im} a_J^i) -\frac{1}{2} \right]^2 \le \frac{1}{4} 
  \eeq
At tree-level, the bound is generally relaxed to
\begin{align}
\mid\mathrm{Re} (a_J^i)\mid \le \frac{1}{2}.
  \end{align}

We assume the external masses of $s_{1,2,3,4}$ are vanishing at high energy limit, and focus on the $J=0$ partial wave.
The modified zeroth partial wave for $s_1s_2\to s_3s_4$ is 
\beq\label{uni-hel}
    a_0 \simeq \frac{1}{16 \pi } \left(2^{-\frac{1}{2}(\delta_{12}+ \delta_{34})} Q_{1234}\right),
\eeq
where $Q_{1234}$ is quartic coupling of $s_1s_2s_3s_4$.

Now we study the unitarity constraints on the 2HDM scalar potential. 
For the scalar potential in Eq. (\ref{V2HDMii}), one can take the uncoupled sets of scalar pairs
\begin{eqnarray} &
\big\{\phi_1^+\phi_2^-, \phi_1^-\phi_2^+, \phi_1 \phi_2, \phi_1 a_2, a_1 \phi_2, a_1 a_2\big\} ,\\
&\big\{ \phi_1^+ \phi_1, \phi_1^+ a_1, \phi_2^+ \phi_2, \phi_2^+ a_2 \big\} ,\\
&\big\{ \phi_1^+ \phi_2, \phi_1^+ a_2, \phi_2^+ \phi_1, \phi_2^+ a_1 \big\} ,\\
&\big\{ \phi_1 a_1, \phi_2 a_2 \big\} ,\\
&\Big\{ \phi_1^+ \phi_1^-, \phi_2^+ \phi_2^-, \phi_1 \phi_1, \phi_2 \phi_2,
a_1 a_1, a_2 a_2\big\} 
\end{eqnarray}
to construct the matrix containing the tree-level amplitudes for $s_1s_2\to s_3s_4$.
We can obtain different eigenvalues of these matrices \cite{unit-2h1,unit-2h2}
\begin{eqnarray}
a_\pm^{} &=& \tfrac{3}{2}(\lambda_1+\lambda_2) \pm \sqrt{\tfrac{9}{4}(\lambda_1-\lambda_2)\raisebox{0.3pt}{$^2$}+(2\lambda_3+\lambda_4)^2} \,, \\
b_\pm^{} &=& \tfrac{1}{2}(\lambda_1+\lambda_2) \pm
\sqrt{\tfrac{1}{4}(\lambda_1-\lambda_2)\raisebox{0.3pt}{$^2$}+\lambda_4^2} \,, \\
c_\pm^{} \,&=&\, \tfrac{1}{2}(\lambda_1+\lambda_2) \pm
\sqrt{\tfrac{1}{4}(\lambda_1-\lambda_2)\raisebox{0.3pt}{$^2$}+\lambda_5^2} \,, \\
{\tt e}_\pm^{} &=& \lambda_3^{} + 2 \lambda_4^{} \pm 3 \lambda_5^{} \,, \\
{\tt f}_\pm^{} \,&=&\, \lambda_3^{} \pm \lambda_4^{} \,,\\
{\tt g}_\pm \,&=&\, \lambda_3^{} \pm \lambda_5^{} \,.
\end{eqnarray}
The unitarity of the scattering process $s_1^{}s_2^{}\to s_3^{}s_4^{}$ leads to
\begin{eqnarray} \label{unitarity}
|a_{\pm}|, |b_\pm|, |c_\pm|, |{\tt e}_\pm|, |{\tt f}_\pm|, |{\tt g}_\pm|
\,\le\, 8\pi \,.
\end{eqnarray}
Here we stress that the conditions of Eq. (\ref{unitarity}) just indicate the approximate level above which the tree-level scattering amplitudes do not provide reliable results anymore. 
The problem is that we cannot rely on perturbative expansion when analyzing scattering, and therefore Eq. (\ref{unitarity}) is just our safety check, not the strict theory limitation.
In addition, we take the standard approach to derive Eq. (\ref{uni-hel}) and Eq. (\ref{unitarity}) and only consider quartic point-like couplings in the high energy limit. 
 At finite energy, the additional diagrams of $s,~t,~u$ channel in $s_1s_2 \to s_3s_4$ scattering can give some corrections to Eq. (\ref{uni-hel}) and Eq. (\ref{unitarity}) \cite{uni-lowe}.

\subsection{Oblique parameters}
The 2HDM can give additional contributions to gauge boson self-energies
by the exchange of extra Higgs fields in the loops. The oblique parameters $S$, $T$ and $U$ 
were used to describe deviations of 2HDM from the SM, which are given as \cite{stu1,stu2,stu3}
\begin{eqnarray} 
 S\; &=& \; \frac{1}{\pi M_Z^2}\,  \Biggl\{ \sin^2(\beta-\alpha)\; \biggl[
\mathcal{B}_{22}(M_Z^2;M_Z^2,M_h^2)- M_Z^2\,
\mathcal{B}_{0}(M_Z^2;M_Z^2,M_h^2) + \mathcal{B}_{22}(M_Z^2;M_H^2,M_A^2)  \biggr] \nonumber \\
&& +\; \cos^2(\beta-\alpha)\; \biggl[
\mathcal{B}_{22}(M_Z^2;M_Z^2,M_H^2)- M_Z^2\,
\mathcal{B}_{0}(M_Z^2;M_Z^2,M_H^2) + \mathcal{B}_{22}(M_Z^2;M_h^2,M_A^2)  \biggr] \nonumber \\
&& -\; \mathcal{B}_{22}(M_Z^2;{M^2_{H^\pm}},{M^2_{H^\pm}})
-\mathcal{B}_{22}(M_Z^2;M_Z^2,M_{h,\mathrm{ref}}^2)
+ M_Z^2\,\mathcal{B}_{0}(M_Z^2;M_Z^2,M_{h,\mathrm{ref}}^2)\Biggr\}\, ,\\ 
T\;  &=&\;  \frac{1}{16\pi M_W^2s_W^2}\,  \Biggl\{\sin^2(\beta-\alpha)\; \biggl[
 \mathcal{F}(M_{H^\pm}^2,M_H^2) - \mathcal{F}(M_H^2,M_A^2) + 3  \!\  \mathcal{F}(M_Z^2,M_h^2) - 3 \!\ \mathcal{F}(M_W^2,M_h^2)  \biggr] \nonumber \\
&& + \; \cos^2(\beta-\alpha)\; \biggl[
 \mathcal{F}(M_{H^\pm}^2,M_h^2) - \mathcal{F}(M_h^2,M_A^2) + 3  \!\  \mathcal{F}(M_Z^2,M_H^2) - 3 \!\ \mathcal{F}(M_W^2,M_H^2)  \biggr] \nonumber \\
&& +\; \mathcal{F}(M_{H^\pm}^2,M_A^2) -  3 \!\ \mathcal{F}(M_Z^2,M_{h,\mathrm{ref}}^2) +3  \!\  \mathcal{F}(M_W^2,M_{h,\mathrm{ref}}^2) \Biggr\}\, 
\end{eqnarray}  
\begin{eqnarray}  
U\; &=&\; \mathcal{H}(M_W^2)-\mathcal{H}(M_Z^2)\nonumber\\
&& +\frac{1}{\pi M_W^2}\, \Biggl\{
  \cos^2(\beta-\alpha) \; \mathcal{B}_{22}(M_W^2;M^2_{H^\pm},M_h^2)+
\sin^2(\beta-\alpha) \; \mathcal{B}_{22}(M_W^2;M^2_{H^\pm},M_H^2) \notag \\  
&& +\;  \mathcal{B}_{22}(M_W^2;M^2_{H^\pm},M_A^2)  - 2  \!\ \mathcal{B}_{22}(M_W^2;M^2_{H^\pm},M^2_{H^\pm}) 
   \Biggr\} \notag  \\
&& -\frac{1}{\pi M_Z^2}\, \Biggl\{
 \cos^2(\beta-\alpha)  \; \mathcal{B}_{22}(M_Z^2;M^2_h,M_A^2)  +
\sin^2(\beta-\alpha)  \; \mathcal{B}_{22}(M_Z^2;M^2_H,M_A^2) \notag  \\
&& -\;  \mathcal{B}_{22}(M_Z^2;M_{H^\pm}^2,M_{H^\pm}^2) \Biggr\} \, ,
\end{eqnarray} 
where
\begin{eqnarray}
\mathcal{H}(M_V^2)\;&\equiv& \;  \frac{1}{\pi M_V^2}\; \Biggl\{  \sin^2(\beta-\alpha) \;\biggl[
\mathcal{B}_{22}(M_V^2; M_V^2,M_h^2)
-M_V^2\,\mathcal{B}_0(M_V^2;M_V^2,M_h^2)\biggr]
\notag \\
&& +\; \cos^2(\beta-\alpha) \;\biggl[
\mathcal{B}_{22}(M_V^2; M_V^2,M_H^2)
-M_V^2\,\mathcal{B}_0(M_V^2;M_V^2,M_H^2)\biggr]
\notag \\
&& -\; \mathcal{B}_{22}(M_V^2; M_V^2,M_{h,\mathrm{ref}}^2)
+M_V^2\,\mathcal{B}_0(M_V^2;M_V^2,M_{h,\mathrm{ref}}^2)\Biggr\}\, .
\end{eqnarray}

\begin{figure}[tb]
 \epsfig{file=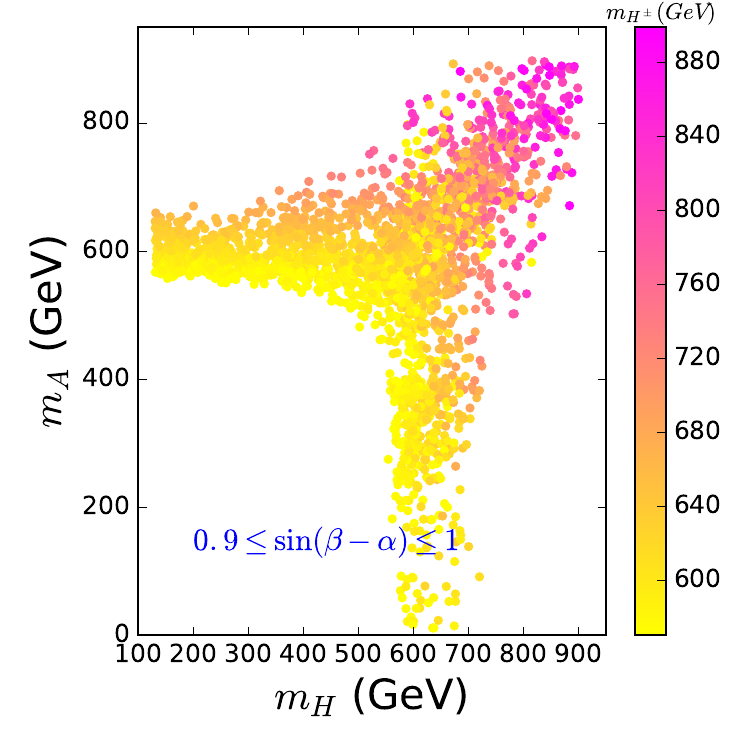,height=7.0cm}
 \epsfig{file=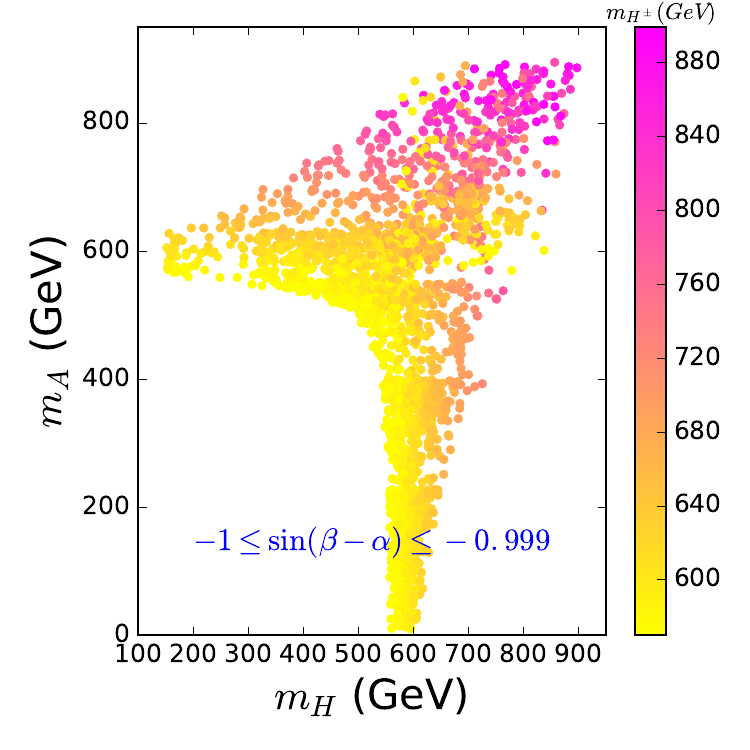,height=7.0cm}
\vspace{-0.5cm} 
\caption{Scatter plots of $m_A$ and $m_H$ satisfying the constraints of vacuum stability, unitarity, 
perturbativity, and the oblique parameters for 570 GeV $\leq m_{H^{\pm}}\leq$ 900 GeV, taken 
from \cite{cpc-wr}.} \label{stu}
\end{figure}

The loop functions are given by
\bea
\label{eqnb}
B_{22}(q^2;m_1^2,m_2^2) &=& \frac{1}{4}\, (\Delta+1)\, [m_1^2+m_2^2-\frac{1}{3}\, q^2]-\frac{1}{2}\,\int^1_0 dx\; X\; \log{(X-i\epsilon)}\, ,\nonumber
\\
B_{0}(q^2;m_1^2,m_2^2) &=& \Delta-\int^1_0 dx\; \log{(X-i\epsilon)}\, ,\nonumber
\\
\mathcal{F}(m_1^2,m_2^2) &=& \frac{1}{2}\, (m_1^2+m_2^2)-\frac{m_1^2m_2^2}{m_1^2-m_2^2}\;
\log{\left(\frac{m_1^2}{m_2^2}\right)}\, ,
\eea
where
\begin{equation}
X \;\equiv\; m_1^2\, x + m_2^2\, (1-x) -q^2\, x(1-x)\, ,
\qquad\quad
\Delta \;\equiv\; \frac{2}{4-d}+\ln 4\pi-\gamma_E\, ,
\end{equation}
in $d$ space-time dimensions. The $\mathcal{B}_{22}$ and $\mathcal{B}_{0}$ functions are defined as
\bea
\mathcal{B}_{22}(q^2;m_1^2,m_2^2) &\equiv&
B_{22}(q^2;m_1^2,m_2^2)-B_{22}(0;m_1^2,m_2^2)\,,\label{b22} \\[6pt]
\mathcal{B}_{0}(q^2;m_1^2,m_2^2) &\equiv& B_{0}(q^2;m_1^2,m_2^2)-B_{0}(0;m_1^2,m_2^2)\,.\label{bzero}
\eea

The above expressions show that the oblique parameters $S$, $T$ and $U$ are sensitive to the mass splitting of extra Higgs bosons.
If $h$ is taken as the 125 GeV Higgs, $H$ or $A$ is favored to have small mass splitting from $H^\pm$.
Fig. \ref{stu} shows $m_H$ and $m_A$ for type-II 2HDM allowed by the global fit values to the oblique parameters \cite{pdg2018},
\beq
S=0.02\pm 0.10,~~  T=0.07\pm 0.12,~~ U=0.00 \pm 0.09, 
\eeq
with correlation coefficients
\beq
\rho_{ST} = 0.89,~~  \rho_{SU} = -0.54,~~  \rho_{TU} = -0.83.
\eeq
In Fig. \ref{stu} $m_{H^\pm} > 570$ GeV is taken to satisfy the constraints of the experimental data of $b\to s\gamma$ \cite{bsr570}.

Very recently the CDF collaboration reported their new result for the $W$-boson mass measurement \cite{cdfmw}
\bea
 m_W=80.4335 \pm 0.0094 {\rm GeV},
\eea 
which has an
approximate $7\sigma$ deviation from the SM prediction, $m_W$(\rm SM)=$80.357 \pm 0.006$  GeV \cite{smmw}.
The shifted $W$-mass modifies the global fit values to $S$, $T$, and $U$ \cite{2204.03796}
\beq\label{fit-stu}
S=0.06\pm 0.10, ~~T=0.11\pm 0.12,~~U=0.14 \pm 0.09,
\eeq
with correlation coefficients
\beq
\rho_{ST} = 0.9, ~~\rho_{SU} = -0.59, ~~\rho_{TU} = -0.85.
\eeq
The $W$-boson mass can be inferred from the following relation \cite{w-stu},
\beq
m_W^2=m_W^2 {\rm (SM)}+\frac{\alpha c_W^2}{c_W^2-s_W^2}m_Z^2 (-\frac{1}{2}S+c_W^2T+\frac{c_W^2-s_W^2}{4s_W^2}U).
\eeq

In the 2HDM, the correction to $T$ is usually larger than $S$ and $U$. In order to accommodate the $W$-mass reported by the CDF II collaboration,
the 2HDM needs to give an appropriate value of $T$. Therefore, $H/A$ is disfavored to degenerate in mass with $H^\pm$.
Various types of 2HDMs have been used to explain the $W$-mass \cite{mw2h-1,mw2h-2,mw2h-3,mw2h-4,mw2h-5,mw2h-6,mw2h-7,mw2h-8,mw2h-9,mw2h-10,mw2h-11,mw2h-12,mw2h-13,mw2h-14}.
Ref. \cite{mw2h-4} discussed the CDF $W$-mass in the 2HDM with an exact $Z_4$ symmetry and found that 
the CDF $W$-mass favors the mass splitting between $H^\pm$ and $H/A$ to be larger than 10 GeV,
and allows $H$ and $A$ to degenerate.
The $m_{H^\pm}$ and $m_{A}$ are favored to be smaller than 650 GeV for $m_H<120$ GeV, and allowed to
have more large values with increasing of $m_H$.

\section{Constraints from LHC searches for Higgs bosons}
\subsection{Signal data of the 125 GeV Higgs}
In the four types of 2HDMs, the neutral Higgs Yukawa couplings normalized to the SM are given by
\bea\label{hffcoupling} &&
y_{h}^{f_i}=\left[\sin(\beta-\alpha)+\cos(\beta-\alpha)\kappa_f\right], \\
&&y_{H}^{f_i}=\left[\cos(\beta-\alpha)-\sin(\beta-\alpha)\kappa_f\right], \\
&&y_{A}^{f_i}=-i\kappa_f~{\rm (for~u)},\\
&& y_{A}^{f_i}=i \kappa_f~{\rm (for~d,~\ell)}.
\eea
The neutral Higgs couplings with gauge bosons normalized to the
SM are
\beq
y^{V}_h=\sin(\beta-\alpha),~~~
y^{V}_H=\cos(\beta-\alpha),\label{hvvcoupling}
\eeq
with $V$ denoting $W$ or $Z$.

The analyses of ATLAS and CMS collaborations show that the coupling strengths of the
discovered 125 GeV boson agree well with the SM Higgs boson, but the sign of 
the couplings cannot measured directly. If we take $h$ as the 125 GeV Higgs boson,
its couplings have two different cases:
\bea
&&y_h^{f_i}~\times~y^{V}_h > 0~~~({\rm for~SM-like~couplings}),\\
&&y_h^{f_i}~\times~y^{V}_h < 0~~~({\rm for~wrong-sign~Yukawa~couplings}).\label{wrongsign}
\eea
In the case of the SM-like couplings, the couplings of the 125 GeV Higgs are very close to those in the SM , which has an alignment limit. 
In the exact alignment limit \cite{align-1,align-2}, namely $\cos(\beta-\alpha) = 0$,
from Eq.  (\ref{hffcoupling}) and Eq. (\ref{hvvcoupling}) we see that $h$ has the same couplings 
to the fermions and gauge
bosons as in the SM, and the heavy CP-even Higgs $H$ has no couplings to the gauge bosons.

Now we discuss the wrong-sign Yukawa couplings \cite{cpc-wr,ws-1,ws-2,ws-3,ws-4,ws-4-1,ws-5,ws-6,ws-7,ws-8,
ws-9,ws-10,ws-10-1,ws-10-2,ws-11,ws-12,ws-13}. The signal data of the 125 GeV Higgs require the absolute values of 
$y_h^{f_i}$ and $y^{V}_h$ to be close to 1.0. Thus, we approximately express $y_h^{f_i}$ and $y^{V}_h$ 
with $\epsilon$ and $\cos(\beta-\alpha)$ as 
\begin{align}  &y_h^{f_i}=-1+\epsilon,~~y^{V}_h\simeq 1-0.5\cos^2(\beta-\alpha) ~~{\rm for}~ \sin(\beta-\alpha) >0~{\rm and}~\cos(\beta-\alpha) >0~,\\
& y_h^{f_i}=1-\epsilon,~~y^{V}_h\simeq -1+0.5\cos^2(\beta-\alpha) ~~{\rm for}~ \sin(\beta-\alpha)<0~{\rm and}~\cos(\beta-\alpha) >0. \end{align}
\begin{figure}[tb]
\begin{center}
 \epsfig{file=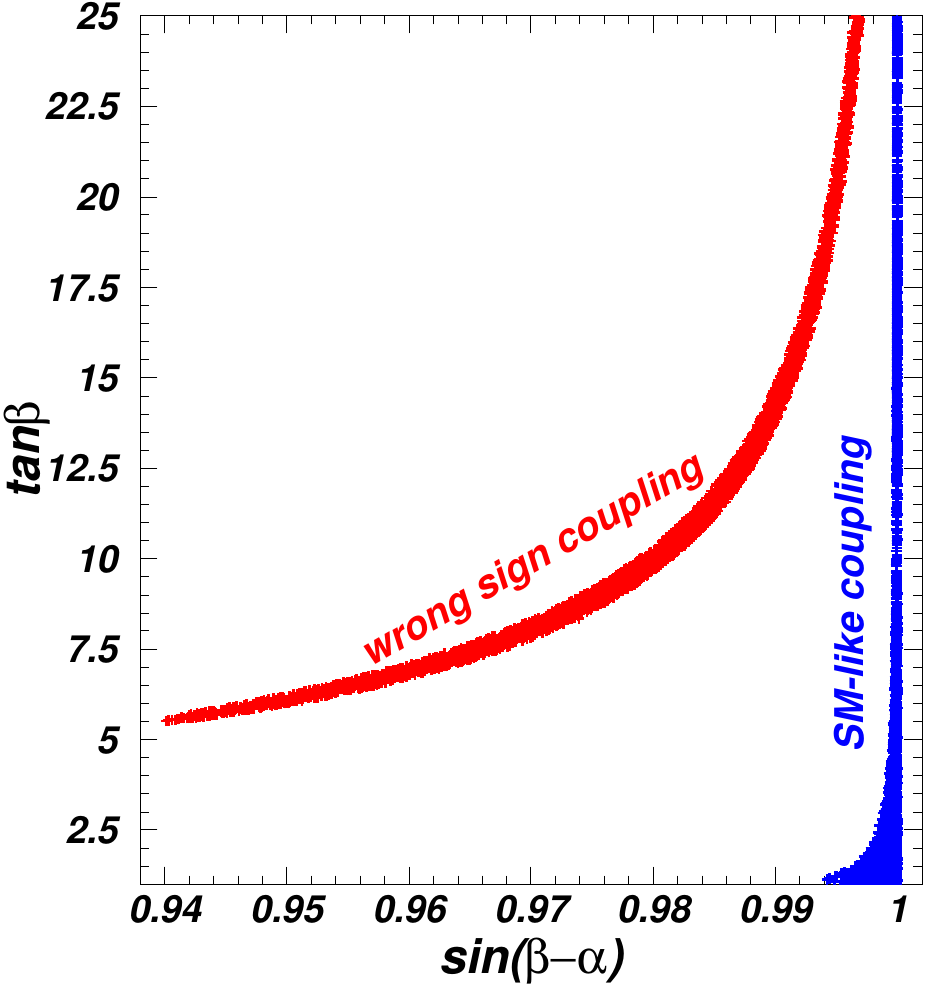,height=7.0cm}
 \epsfig{file=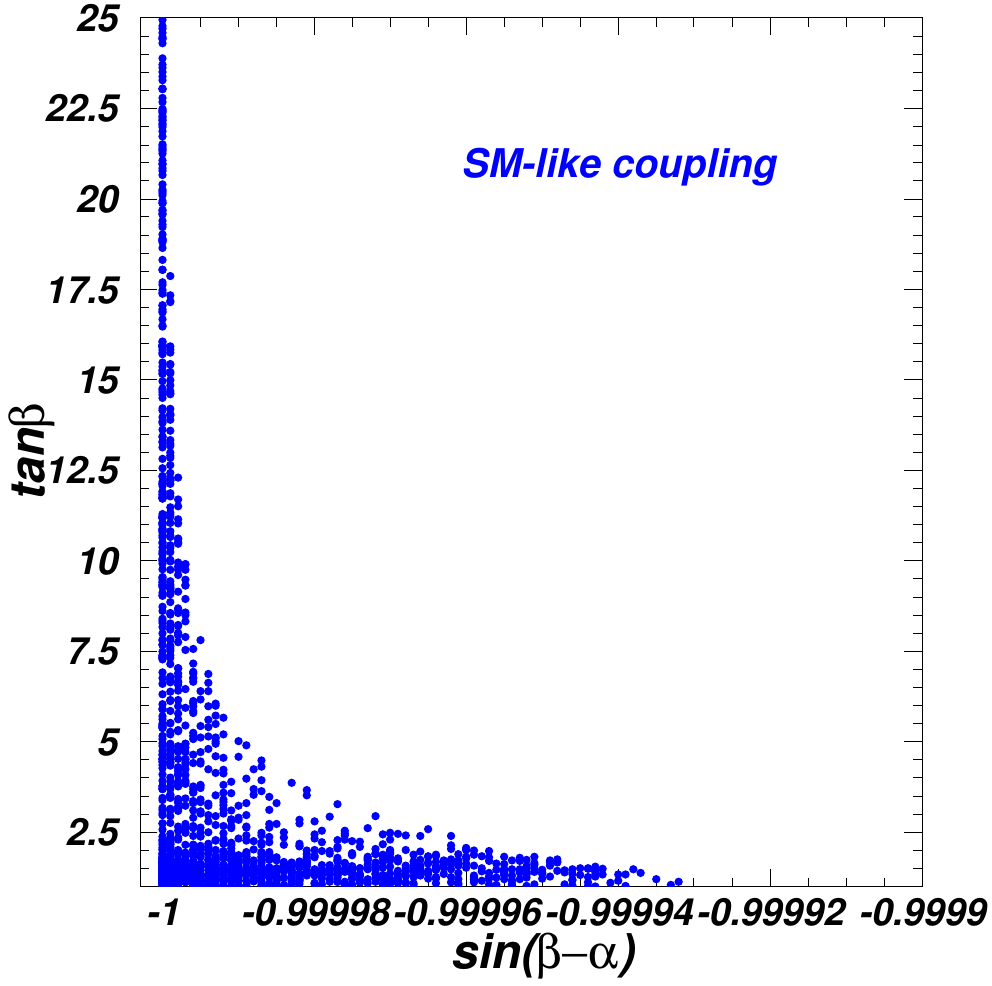,height=7.0cm}
\end{center}
\vspace{-0.3cm} 
\caption{Scatter plots of $\sin(\beta-\alpha)$ and $\tan\beta$ of type-II model 
satisfying the constraints of
the 125 GeV Higgs signal data, taken from \cite{2010.03730}.} \label{125higgs}
\end{figure}

From Eq. (\ref{hffcoupling}) we can get
\begin{align}\label{wrcp}
&\kappa_f=\frac{-2+\varepsilon+0.5\cos(\beta-\alpha)^2}{\cos(\beta-\alpha)}<<-1 ~{\rm for}~ \sin(\beta-\alpha) >0~{\rm and}~\cos(\beta-\alpha) >0~,\\
&\kappa_f=\frac{2-\varepsilon-0.5\cos(\beta-\alpha)^2}{\cos(\beta-\alpha)} >>1 ~{\rm for}~ \sin(\beta-\alpha) <0~{\rm and}~\cos(\beta-\alpha) >0~.
\end{align}
In the four types of 2HDMs, the measurement of the branching fraction of $b\to s\gamma$ favors 
a $\tan\beta$ greater than 1.
Therefore, for $\sin(\beta-\alpha) > 0$ and $\cos(\beta-\alpha) >0$, there  may exist wrong-sign 
Yukawa couplings for the down-type quarks and leptons in the type-II model, for the leptons in the L2HDM, 
and for the down-type quarks in the flipped 2HDM.

Fig. \ref{125higgs} shows $\sin(\beta-\alpha)$ and $\tan\beta$ of type-II model allowed by 
the 125 GeV Higgs signal data. The value of $\sin(\beta-\alpha)$ in the case 
of the wrong-sign Yukawa couplings 
is allowed to deviate from 1 more sizably than in the case of the SM-like couplings. 
In the case of the wrong-sign Yukawa couplings, $\tan\beta$ has stringent upper and lower bounds 
for a given value of $\sin(\beta-\alpha)$.

\subsection{Searches for additional scalars at LHC}
\begin{table}
\caption{The upper limits at 95\%  C.L. on the production cross section times branching ratio 
for the channels of $H$ and $A$ searches at the LHC.}
\label{tablhc}
\vspace{0.3cm} 
\begin{footnotesize}
\begin{tabular}{| c | c | c | c |}
\hline
\textbf{Channel} & \textbf{Experiment} & \textbf{Mass range [GeV]}  &  \textbf{Luminosity} \\
\hline
{$gg/b\bar{b}\to H/A \to \tau^{+}\tau^{-}$} & CMS 13 TeV \cite{1709.07242}& 200-2250   & 36.1 fb$^{-1}$ \\
{$gg/b\bar{b}\to H/A \to \tau^{+}\tau^{-}$} & ATLAS 13 TeV \cite{2002.12223}& 200-2500   & 139 fb$^{-1}$ \\
\hline
{$gg\to H/A \to t\bar{t}$} & CMS 13 TeV \cite{1908.01115}& 400-750   & 35.9 fb$^{-1}$ \\
\hline
{$gg\to H/A \to \gamma\gamma$~+~$t\bar{t}H/A~(H/A\to \gamma\gamma)$}& CMS 13 TeV \cite{HIG-17-013-pas}& 70-110 & 35.9 fb$^{-1}$ \\
{$VV\to H \to \gamma\gamma$~+~$VH~(H\to \gamma\gamma)$}& CMS 13 TeV \cite{HIG-17-013-pas}& 70-110 & 35.9 fb$^{-1}$ \\
\hline

{$gg/VV\to H\to W^{+}W^{-}~(\ell\nu qq)$} & ATLAS 13 TeV  \cite{1710.07235}& 200-3000  &  36.1 fb$^{-1}$\\
{$gg/VV\to H\to W^{+}W^{-}~(e\nu \mu\nu)$} & ATLAS 13 TeV  \cite{1710.01123}& 200-3000  &  36.1 fb$^{-1}$\\
{$gg/VV\to H\to W^{+}W^{-}$} & CMS 13 TeV  \cite{1912.01594}& 200-3000  &  35.9 fb$^{-1}$\\
\hline
$gg/VV\to H\to ZZ $ & ATLAS 13 TeV~\cite{1712.06386} & 200-2000 & 36.1 fb$^{-1}$ \\
$gg/VV\to H\to ZZ $ & ATLAS 13 TeV~\cite{1708.09638} & 300-5000 & 36.1 fb$^{-1}$ \\
$gg/VV\to H\to ZZ $ & ATLAS 13 TeV~\cite{2009.14791} & 200-2000 & 139 fb$^{-1}$ \\

\hline
$gg \to H\to hh \to b\bar{b}b\bar{b}$ & CMS 13 TeV~\cite{1710.04960} & 750-3000  &  35.9 fb$^{-1}$ \\
$gg \to H\to hh \to (b\bar{b}) (\tau^{+}\tau^{-})$ & CMS 13 TeV~\cite{1707.02909} & 250-900  &  35.9 fb$^{-1}$ \\
$pp \to H\to hh $ & CMS 13 TeV~\cite{1811.09689} & 250-3000  &  35.9 fb$^{-1}$ \\
$gg \to H\to hh \to b\bar{b}ZZ$ & CMS 13 TeV~\cite{2006.06391} & 260-1000  &  35.9 fb$^{-1}$ \\

$gg \to H\to hh \to b\bar{b}\tau^{+}\tau^{-}$ & CMS 13 TeV~\cite{2007.14811} & 1000-3000  &  139 fb$^{-1}$ \\
\hline


{$gg/b\bar{b}\to A\to hZ\to (b\bar{b})Z$}& ATLAS 13 TeV \cite{1712.06518}& 200-2000 & 36.1 fb$^{-1}$  \\

{$gg/b\bar{b}\to A\to hZ\to (b\bar{b})Z$}& CMS 13 TeV \cite{1903.00941}& 225-1000 & 35.9 fb$^{-1}$  \\

{$gg\to A\to hZ\to (\tau^{+}\tau^{-}) (\ell \ell)$}& CMS 13 TeV \cite{1910.11634}& 220-400 & 35.9 fb$^{-1}$  \\
\hline

$gg/b\bar{b}\to A(H)\to H(A)Z\to (b\bar{b}) (\ell \ell)$ & ATLAS 13 TeV \cite{1804.01126}& 130-800 & 36.1 fb$^{-1}$ \\

$gg\to A(H)\to H(A)Z\to (b\bar{b}) (\ell \ell)$ & CMS 13 TeV \cite{1911.03781}& 30-1000 & 35.9 fb$^{-1}$ \\
\hline
\end{tabular}
\end{footnotesize}
\end{table}

The ATLAS and CMS collaborations have searched for an additional scalar from its decay into 
various SM channels 
or from its exotic decays. Since the Yukawa couplings of down-type quarks and leptons can be both
enhanced by a factor of $\tan\beta$, the type-II model can be more stringently  constrained 
than other three types of models by the flavor observables and the LHC searches for additional
Higgs.

At the LHC, the dominant production processes of $H$ and $A$ are from the gluon-gluon fusions, 
which are generated by exchanging top quark and $b$-quark in the loops.
There may be destructive interference between contributions of $b$-quark
loop and top quark loop. The $\textsf{SusHi}$ \cite{sushi} was used to calculate the cross sections for $H$ and $A$ in the
gluon fusion and $b\bar{b}$-associated production at NNLO in QCD, while the 
$\textsf{2HDMC}$ was employed to precisely calculate the branching ratios of the
various decay modes of $H$ and $A$ \cite{2hc-1}. 
\begin{figure}[tb]
\begin{center}
 \epsfig{file=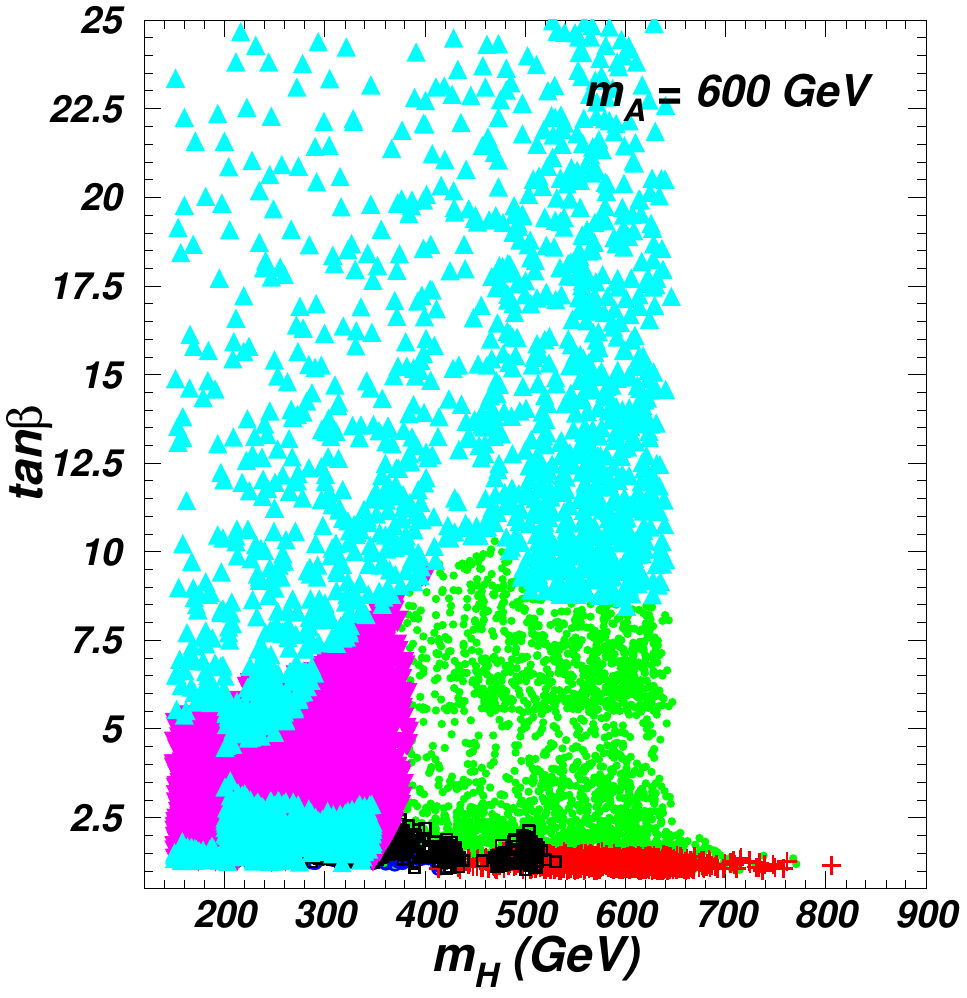,height=7.0cm}
 \epsfig{file=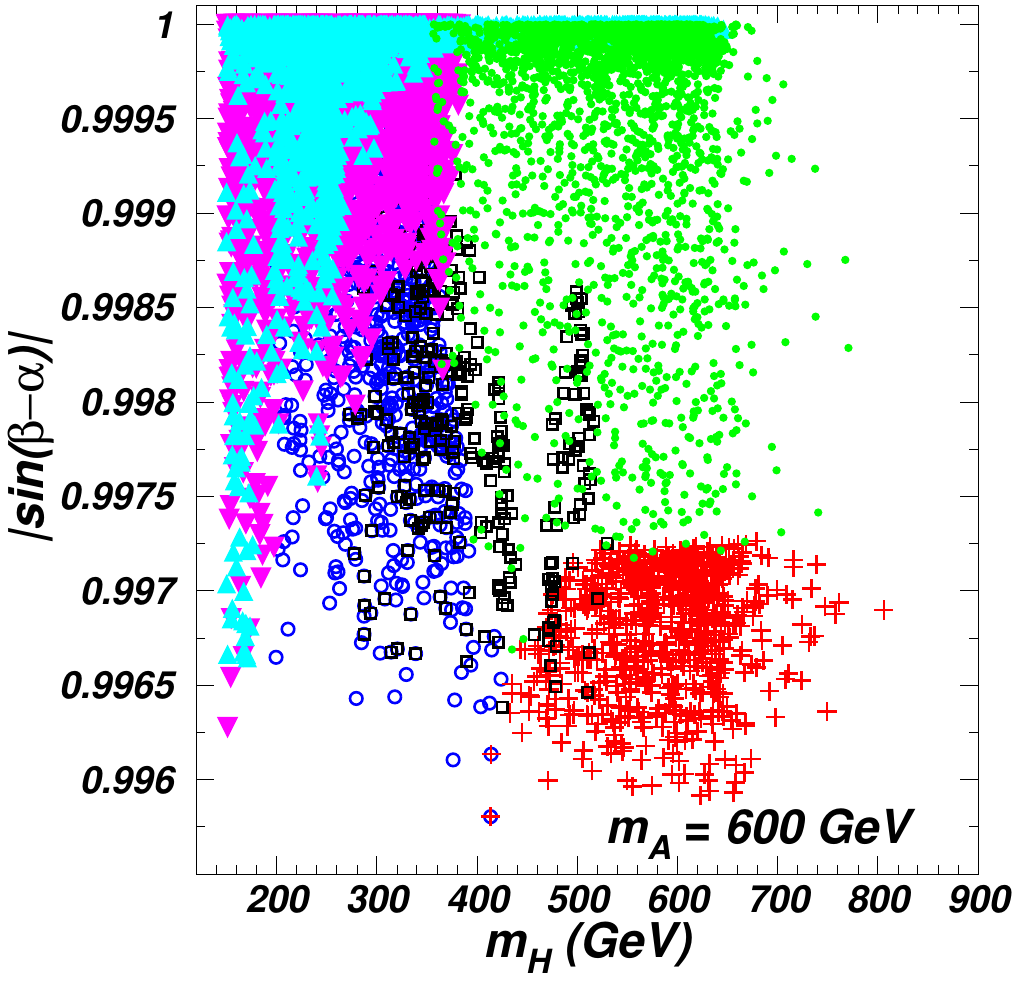,height=7.0cm}
 \end{center}
\vspace{-0.5cm} 
\caption{The surviving samples with the SM-like couplings of type-II model, taken from \cite{2010.03730}.
 The triangles (sky blue), circles (royal blue), squares (black), inverted triangles (purple), 
and pluses (red) are respectively excluded by the $H/A\to \tau^+ \tau^-$, $H\to
WW,~ZZ,\gamma\gamma$, $H\to hh$, $A\to HZ$, and $A\to hZ$ channels at the LHC. The bullets (green) samples are
allowed by various LHC direct searches.} \label{lhcsm}
\end{figure}

\begin{figure}[tb]
\begin{center}
 \epsfig{file=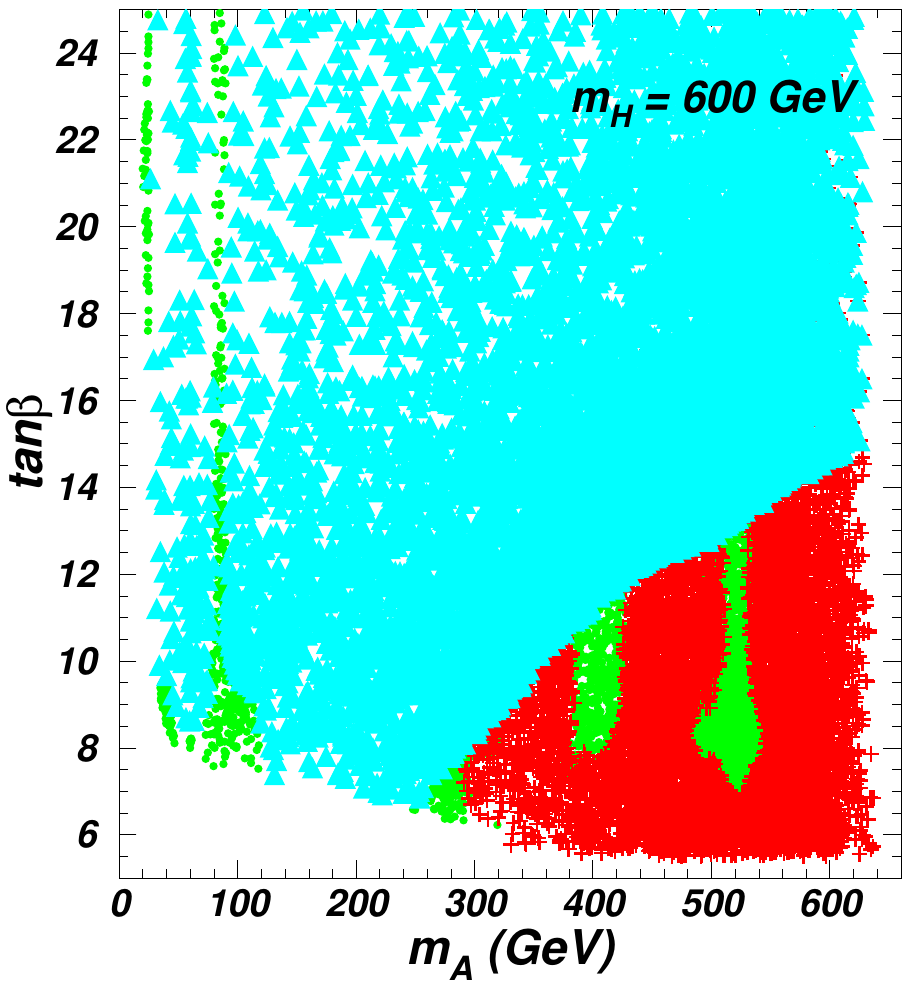,height=7.0cm}
 \epsfig{file=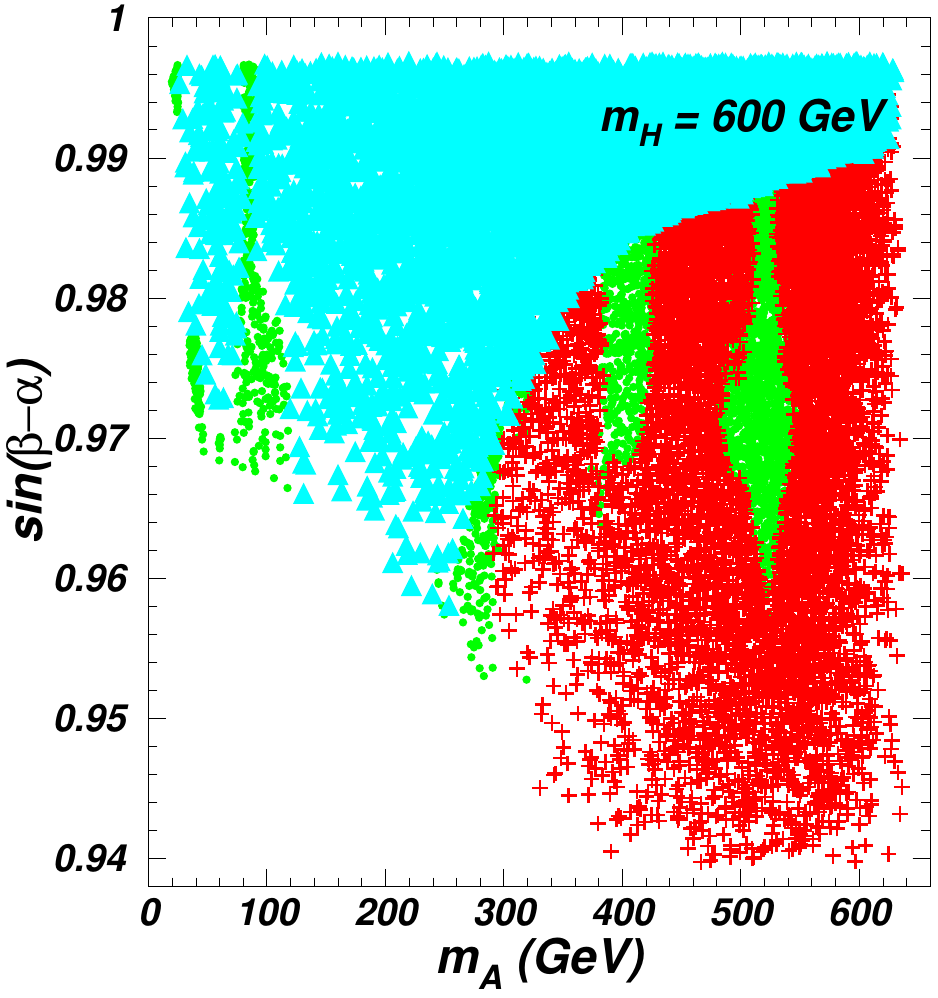,height=7.0cm}
 \epsfig{file=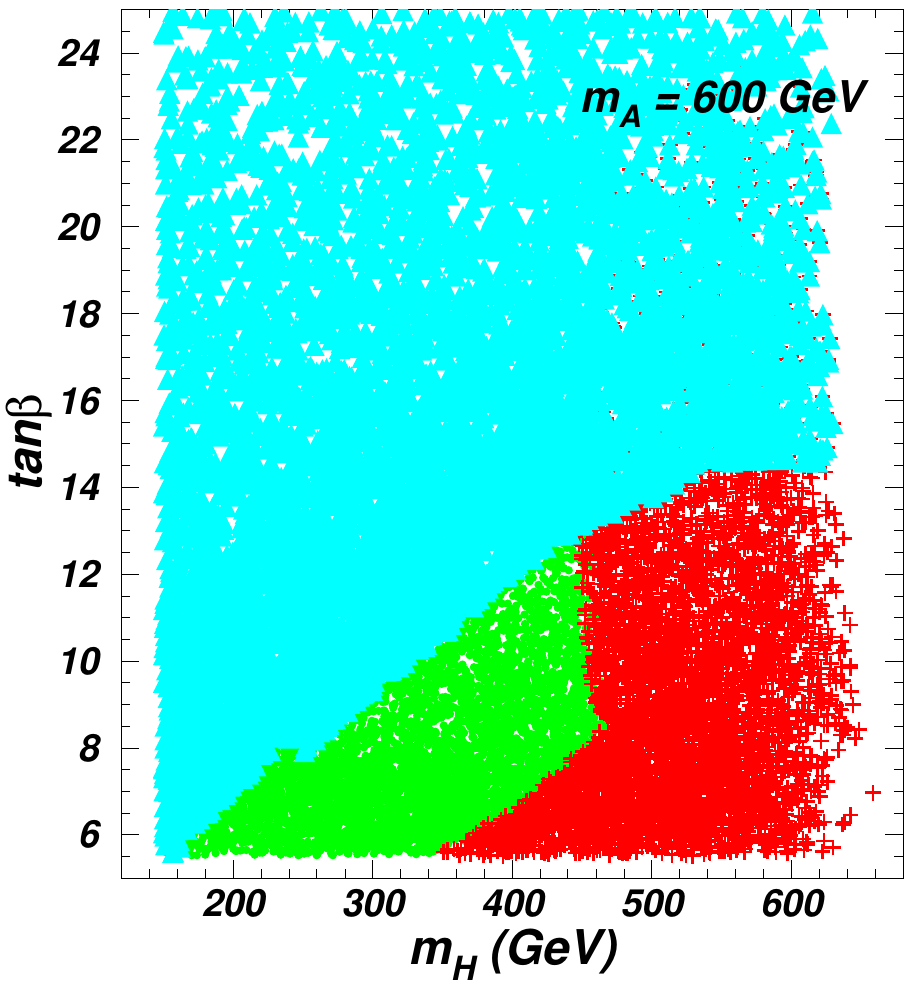,height=7.0cm}
 \epsfig{file=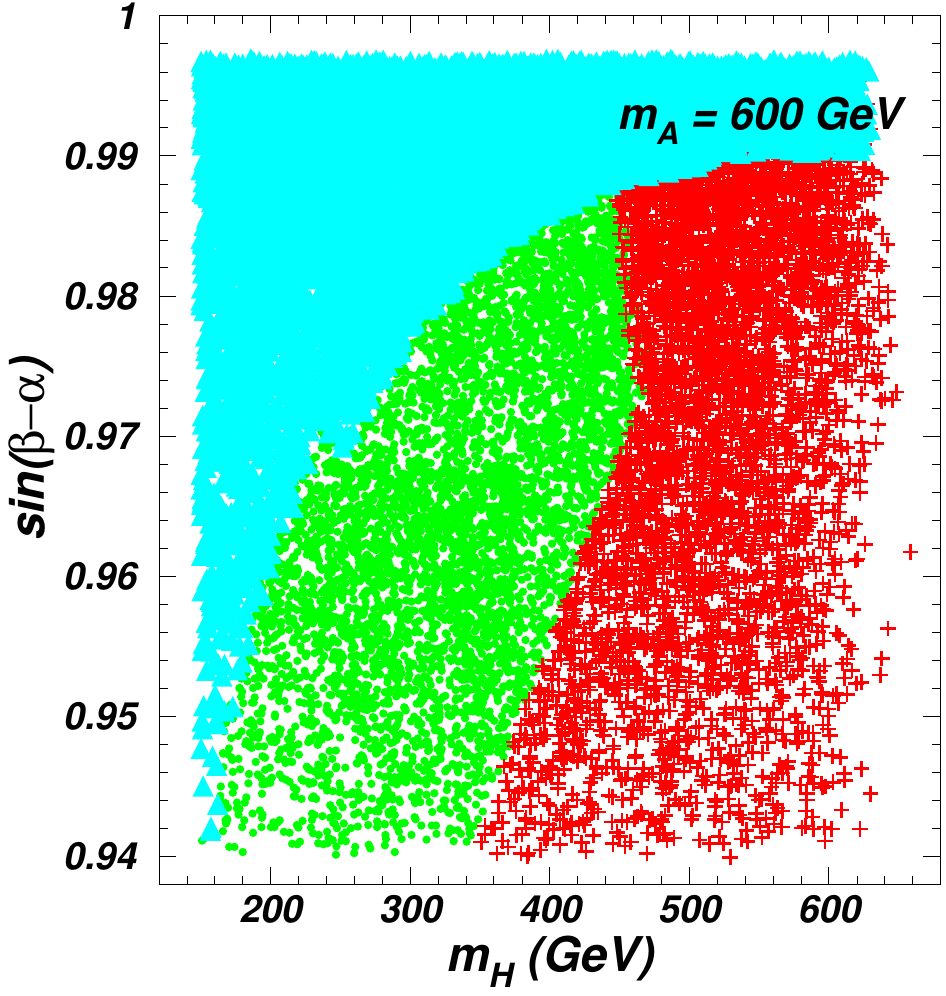,height=7.0cm}
 \end{center}
\vspace{-0.5cm} 
\caption{The surviving samples with the wrong-sign Yukawa couplings of type-II model, taken from \cite{cpc-wr}. The triangles (sky blue) 
and pluses (red) are respectively excluded by the $A/H\to \tau^+ \tau^-$ and $A\to hZ$ channels at the LHC.
The bullets (green) are allowed by various LHC direct searches.} \label{lhcwr}
\end{figure}

The studies in \cite{2010.03730,cpc-wr} used a large number of ATLAS and CMS analyses at 
the 8 TeV and 13 TeV LHC to constrain the type-II 2HDM. 
Table ~\ref{tablhc} lists some analyses at the 13 TeV LHC with more than 35.9 fb$^{-1}$
integrated luminosity data.
Fig. \ref{lhcsm} shows the surviving samples with the SM-like coupling of type-II model satisfying various 
LHC direct searches.  
The couplings of $AhZ$ and $AHZ$ are respectively proportional to $\cos(\beta-\alpha)$ and $\sin(\beta-\alpha)$.
For the case of the SM-like coupling, $\mid\sin(\beta-\alpha)\mid$ is very closed to 1. Therefore, the $A\to hZ$ channel fails to constrain
the parameter space, and the $A\to HZ$ channel can exclude many points in the region of $m_H<$ 360 GeV.
The $H/A \to \tau^+ \tau^-$ channels give upper bound on $\tan\beta$, and 
allow $m_H$ to vary from 150 GeV to 800 GeV for appropriate $\tan\beta$ and $\sin(\beta-\alpha)$.
Fig. \ref{lhcsm} shows the joint constraints of
 $H/A\to\tau^+ \tau^-$, $A\to HZ$, $H\to WW,~ZZ,~\gamma\gamma$, and $H\to hh$ exclude the whole 
region of $m_H<360$ GeV.

The surviving samples with the wrong-sign Yukawa couplings of type-II model are shown in Fig. \ref{lhcwr}.
For the case of the wrong-sign Yukawa couplings, the signal data of the 125 GeV Higgs requires 
$\tan\beta>5$ and allows $\sin(\beta-\alpha)$ to be as low as 0.94, as shown in Fig. \ref{125higgs}. 
As a result, the cross sections of $H$ and $A$ in the gluon fusion productions
are sizably suppressed, and only $b\bar{b}\to A \to \tau^+ \tau^-$ and $A\to hZ$ channels 
can be used to constrain the parameter space.
Especially for $m_H=$ 600 GeV, the constraints are very stringent, and the allowed samples are mainly 
distributed in several corners.
Many samples with $m_A$ in the ranges of $30\sim 120$ GeV, $240\sim 300$ GeV, $380\sim 430$ GeV, and $480\sim 550$ GeV
 are allowed for appropriate $\tan\beta$ and $\sin(\beta-\alpha)$. 
Also the samples in the regions of $m_A<20$ GeV and 80 GeV $<m_A<90$ GeV are allowed since 
there is no experimental data of $A \to \tau^+ \tau^-$ channel in these ranges.

For the case of $m_A=$ 600 GeV, the constraints of the $b\bar{b}\to A \to \tau^+ \tau^-$ and $A\to hZ$ channels can 
be relatively relaxed. Many samples of 150 GeV $<m_H<$ 470 GeV are allowed and $m_H>$ 470 GeV is excluded.
For a small $m_H$, the $A\to HZ$ decay will open and increase the total width of $A$. As a result, 
the branching ratio of $A\to hZ$ can be sizably suppressed, and weaken the constraints of the $A\to hZ$ channel. 

Compared to the type-II 2HDM, all the Yukawa couplings of $H$, $A$ and $H^\pm$ in the type-I model can be suppressed by
a large $\tan\beta$, which leads that the searches for additional scalars at the LHC and measurements of the flavor observables are easily 
satisfied. Thus, $H$, $A$ and $H^\pm$ are allowed to have broad mass ranges. There are some recent 
studies on the status of type-I and type-II 2HDMs confronted with the direct searches at the LHC, see, e.g., 
\cite{typ1-lhc-1,typ1-lhc-2,typ1-lhc-3,2202.08807,2107.05650,2004.04172,2102.07136,2112.13679,2010.15057,2006.01164,2005.10576}.

\section{Dark matter observables}
\subsection{Inert 2HDM and dark matter}
Because of the exact $Z_2$ symmetry, the lightest neutral
component $H$ or $A$ is stable and may be considered as a DM candidate.
If taking $H$ as the DM, it requires
\beq
\lambda_5 < 0, ~~~\lambda_4-\mid\lambda_5\mid < 0.
\eeq
Flipping the sign of $\lambda_5$, $A$ will be the DM candidate. 
The parameter $\lambda_{345}=\lambda_3+\lambda_4+\lambda_5$ controls
the $hHH$ coupling, which will affect the signal strengths of the 125 GeV Higgs and the DM observables. 

The main possible annihilation channels include $HH\to f\bar{f},~ VV^{(*)},~hh$ and 
various co-annihilations of the inert scalars.
In addition to the constraints from theory and the oblique parameters as well as 
the signal data of the 125 GeV Higgs, the model should also satisfy the precise 
measurements of the $W$ and $Z$ widths, which requires
\beq
m_A+m_H > m_Z,~~2m_{H^\pm} > m_Z,~~m_A+m_{H^\pm} > m_W,~~m_H+m_{H^\pm} > m_W.
\eeq
The null searches at the LEP exclude two regions \cite{inert-lep-1,inert-lep-2},
\bea
&&m_{H^\pm} < 70 {\rm GeV},\\
&&m_H < 80~ {\rm GeV},~~m_A < 100~ {\rm GeV},~~{\rm and }~m_A-m_H > 8 {\rm GeV}. 
\eea

Considering various relevant theoretical and experimental constraints, the allowed DM mass ranges have been  
discussed, see e.g. \cite{inert-dm1,inert-dm2,inert-dm3,inert-dm4,inert-dm5,inert-dm6,inert-dm7,inert-dm8,inert-dm9,inert-dm10,inert-dm11,inert-dm12}. 
Because of the tension between the signal strength of the 125 GeV Higgs and the relic density, $m_H<$ 55 GeV is disfavored.
In the resonance region of $m_H\simeq \frac{m_h}{2}$, the main annihilation channels are $h$-mediated, primarily into $b\bar{b}$ and $WW$ final states.
The correct relic density can be obtained and the relevant constraints can be satisfied. 
In the region up to around 75 GeV,
the $HH$ pair mainly annihilates to $WW^*$ via the processes mediated by $h$ or 
via the quartic couplings. Under the relevant constraints,
the correct relic density can be rendered for 73 GeV $<m_H<$ 75 GeV. 
For 75 GeV $<m_H<$ 160 GeV, the correct relic density requires $\lambda_{345}$
to be large enough to lead to an appropriate cancelation between diagrams of $VV^*$. 
However, such a large $\lambda_{345}$ is excluded by the DM direct detections. 
In the region between 160 GeV and 500 GeV, the annihilation cross section of
$HH\to W^+W^-$ is too large to produce the exact relic density. In the region of $m_H>$ 500 GeV, 
the exact relic density favors small mass splittings among the three inert 
Higgs bosons, roughly $\leq$ 10 GeV. The large mass splittings tend to enhance the 
cross section of $HH$ annihilation into longitudinal $Z$ and $W$ bosons.

\subsection{Wrong-sign Yukawa couplings and isospin-violating interactions between dark matter and nucleons.}

Although the inert 2HDM may provide a DM candidate, but its mass range is stringently constrained. 
Alternatively, a real singlet scalar DM can be added to the 2HDM, and this DM has different properties 
from the DM in inert 2HDM. Especially for the type-II 2HDM, the 125 GeV Higgs may have wrong-sign 
Yukawa couplings with down-type quarks. If such a Higgs acts as the portal between the DM and 
SM sectors, the model can give the isospin-violating interactions between DM and nucleons, 
which can relax the constraints from the DM direct detections.

A real singlet scalar $S$ is introduced to the type-II 2HDM under a $Z'_2$ symmetry 
in which $S\rightarrow -S$. The potential containing the $S$ field is written as
\cite{2hisos-1}
\begin{eqnarray}
\mathcal{V}_{S}&=&{1\over 2}S^2(\kappa_{1}\Phi_1^\dagger \Phi_1
+\kappa_{2}\Phi_2^\dagger \Phi_2)+{m_{0}^2\over
2}S^2+{\lambda_S\over 4!}S^4\label{potent}.
\end{eqnarray} 
The $S$ field has no vev and may serve as a DM candidate.
The DM mass and the cubic interactions with the neutral Higgs bosons are obtained from Eq.
(\ref{potent}),
\begin{eqnarray}
m_S^2&=&m_0^2+\frac{1}{2}\kappa_1
v^2\cos^2\beta+\frac{1}{2}\kappa_2 v^2\sin^2\beta, \\
-\lambda_{h} vS^2h/2&\equiv& -(-\kappa_{1}\sin\alpha\cos\beta+\kappa_{2}\cos\alpha\sin\beta)vS^2h/2, \\
-\lambda_{H} vS^2H/2&\equiv&
-(\kappa_{1}\cos\alpha\cos\beta+\kappa_{2}\sin\alpha\sin\beta)vS^2H/2.
\label{dmcoup}\end{eqnarray}

\begin{figure}[tb]
\begin{center}
\epsfig{file=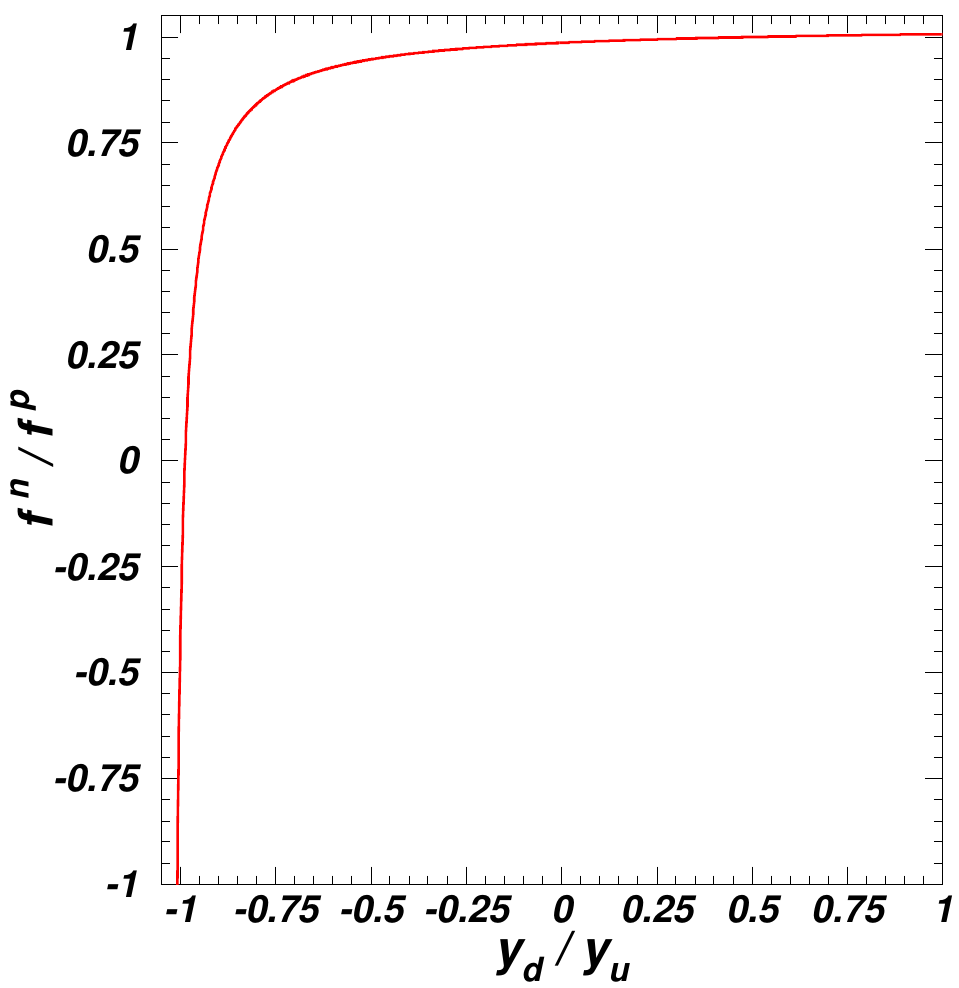,height=7.0cm}
 \epsfig{file=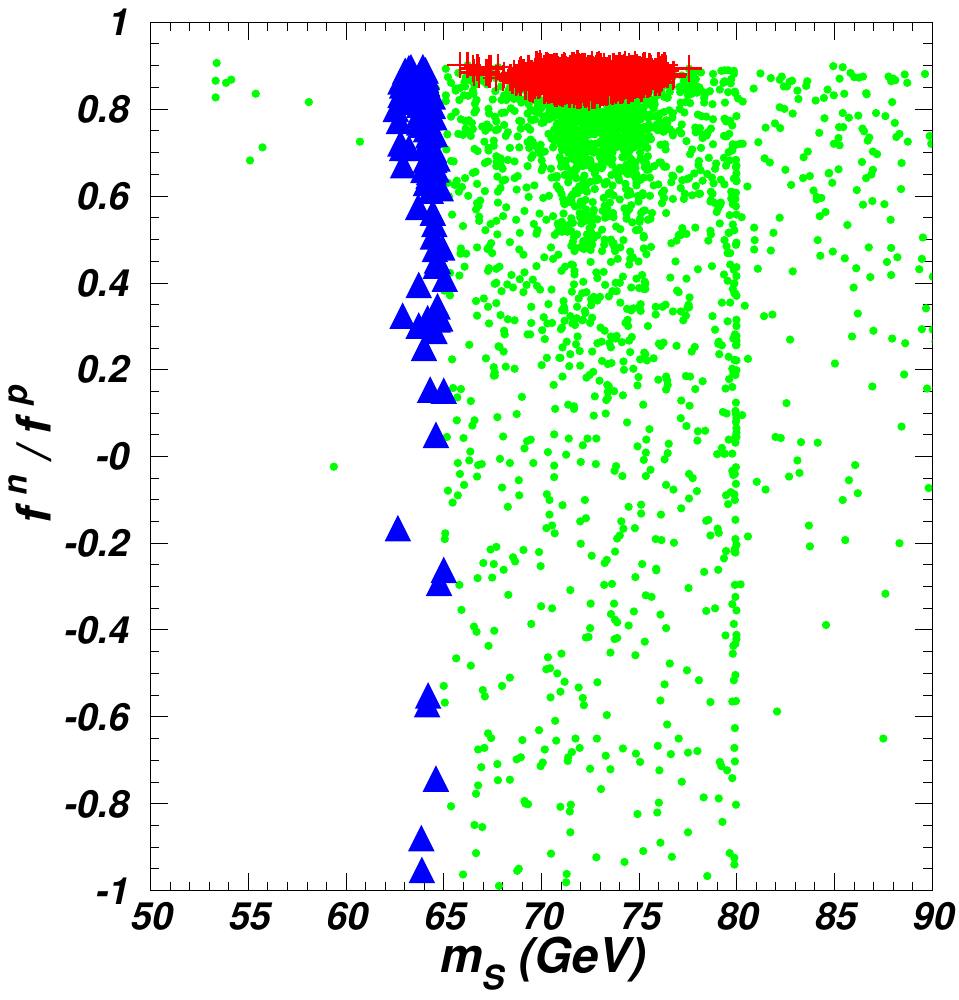,height=7.0cm}
 \end{center}
\vspace{-0.5cm} \caption{Left: $f^n/f^p$ versus $y_d/y_u$ with $y_d$ $(y_u)$ denoting the Yukawa coupling of $hd\bar{d}$ ($hu\bar{u}$) normalized to the SM value \cite{1708.06882}.
 Right: All the samples are allowed by the constraints of the LHC searches and the DM relic density.
The pluses (red) are excluded by the constraints of the spin-independent DM-proton cross section 
from XENON1T (2017), and the triangles (royal blue) are excluded by the Fermi-LAT search for DM annihilation from dSphs \cite{1708.06882}.} \label{dmwr}
\end{figure}
In this model, the elastic scattering of $S$ on a nucleon receives
the contributions from the process with $t$-channel exchange of $h$ and $H$. 
The spin-independent cross section is written as
\cite{sigis},
 \beq \sigma_{p(n)}=\frac{\mu_{p(n)}^{2}}{4\pi m_{S}^{2}}
    \left[f^{p(n)}\right]^{2},
\eeq 
where $\mu_{p(n)}=\frac{m_Sm_{p(n)}}{m_S+m_{p(n)}}$ and  
\beq
f^{p(n)}=\sum_{q=u,d,s}f_{q}^{p(n)}\mathcal{C}_{S
q}\frac{m_{p(n)}}{m_{q}}+\frac{2}{27}f_{g}^{p(n)}\sum_{q=c,b,t}\mathcal{C}_{S
q}\frac{m_{p(n)}}{m_{q}},\label{fpn} 
\eeq 
with $\mathcal{C}_{Sq}=\frac{\lambda_h}{m_h^2} m_q y_q^h + \frac{\lambda_H}{m_H^2} m_q y_q^H$. 
Here $f_{q}^{p}$ ($f_{q}^{n}$) is the form factor at the proton (neutron) for a light quark $q$,
and $f_{g}^{p}$ ($f_{g}^{n}$) is the form factor at the proton (neutron) for gluon \cite{1312.4951},
\begin{eqnarray}
f_{u}^{p}\approx0.0208,\quad & f_{d}^{p}\approx0.0399,\quad &
f_{s}^{p}\approx0.0430,
\quad f_{g}^{p}\approx0.8963,\nonumber\\
f_{u}^{n}\approx0.0188,\quad & f_{d}^{n}\approx0.0440,\quad &
f_{s}^{n}\approx0.0430,\quad  f_{g}^{n}\approx0.8942.
 \label{eq:neuclon-form}
\end{eqnarray}
A simple scenario is to take the 125 GeV Higgs ($h$) as the only portal between the DM and 
SM sectors.
If $f_{q}^{p}\neq f_{q}^{n}$, the $S$-nucleon scattering may be
isospin-violating for the appropriate values of $y_h^d$ and $y_h^u$.

The left panel of Fig. \ref{dmwr} shows that $f^n/f^p$ approaches to 1 with $y_d/y_u$. Namely,  
the $S$-nucleon scattering is isospin-conserving for $y_d=y_u$ and significantly isospin-violating when
$y_d/y_u$ deviates from 1 sizably, especially that there is an opposite sign between $y_d$ and $y_u$. 
The right panel shows that the bounds of the direct detection experiments can be satisfied
in the region $-1<f^n/f^p <$ 0.8. The DM scattering rate with Xe target can be sizably
suppressed for $f^n/f^p\sim -0.7$, which can weaken the constraints from the spin-independent 
DM-nucleon cross section.

There are other DM extensions of 2HDM which accomodate the DM direct detection limits.
In the general 2HDM with a DM, when both $h$ and $H$ are portals between the SM sector and DM, 
and have appropriate couplings, 
the model can achieve the blind spots at DM direct detection, which originates from 
cancellations between interfering diagrams with  
$h$ and $H$ exchanges \cite{blind-dm-1,blind-dm-2}. Besides, in the L2HDM with a DM, 
the quark Yukawa couplings of $H$ can
be significantly suppressed for a very large $\tan\beta$. If such a $H$ field is taken
as the portal between the SM sector and DM, the model can easily weaken the
bound of the DM direct detection and explain the muon $g-2$ \cite{l2hdm-dm-1,2112.15536}.  

\section{Muon anomalous magnetic moment}
\subsection{L2HDM and muon $g-2$}
The muon $g-2$ is a very precisely measured observable and serves as a sensitive probe of 
new physics (for a pedagogical review, see, e.g., \cite{Li:2021bbf}).  
The new Fermilab measurement \cite{fermig2}
combined with E821 data \cite{e821} shows a $4.2\sigma$ deviation from the SM prediction 
\cite{g2qed,g2ew,g2light,g2hvp}.
Such a discrepancy has been explained in various new physics models like the minimal supersymmetry
(see, e.g., \cite{Abdughani:2019wai, Athron:2021iuf, Wang:2021bcx, Endo:2021zal}). 
Among the 2HDMs, the L2HDM can offer an explanation.   

In the L2HDM, the lepton (quark) Yukawa couplings to $H$, $A$ and $H^{\pm}$ can be sizably 
enhanced (suppressed) by a large $\tan\beta$.
The model has been extensively studied to explain the muon $g-2$, 
and the searches at the LHC and low energy precision measurements can
exclude a large part of parameter space for the explanation of muon $g-2$.
The study in \cite{mu2h9} considered the signal data of the 125 GeV Higgs, and found that
the muon $g-2$ explanation favors the 125 GeV Higgs to have wrong-sign Yukawa couplings to the leptons. 
The experimental results of Br$(B_s \to \mu^+ \mu^-)$ can exclude some parameter regions with a very 
light $A$ \cite{mu2h9}. 
Besides, the measurements of lepton flavor universality (LFU) of the $Z$ decays and $\tau$ decays give stringent 
constraints on $\tan\beta$ and 
the mass splittings among $H$, $A$ and $H^\pm$ \cite{mu2h10,mu2h16-2}, and a more precise study was performed in \cite{mu2h16}.
The muon $g-2$ explanation makes the additional Higgs bosons to have $\tau-$rich signatures at 
the LHC, and the study in \cite{mu2h11} first used the chargino/neutralino searches at the 8 TeV LHC 
to constrain the model. The analysis in \cite{mu2h17}
used the constraints of the multi-lepton analyses at the 13 TeV, and found that 
the L2HDM may explain the muon $g-2$ anomaly and produce a strong first order electroweak 
phase transition (SFOEWPT) simultaneously.

In the L2HDM, the additional contributions to the muon $g-2$ are mainly from the
one-loop diagrams and the two-loop Barr-Zee diagrams mediated by $A$, $H$ and $H^\pm$.
The one-loop contributions is given by \cite{mu2h1-1,mu2h1-2,mu2h1-3} 
 \beq
    \Delta a_\mu^{\mbox{$\scriptscriptstyle{\rm 2HDM}$}}({\rm 1-loop}) =
    \frac{G_F \, m_{\mu}^2}{4 \pi^2 \sqrt{2}} \, \sum_j
    \left (y_{\mu}^j \right)^2  r_{\mu}^j \, f_j(r_{\mu}^j),
\label{amuoneloop}
\end{equation}
where $j = H,~ A ,~ H^\pm$, $r_{\mu}^ j =  m_\mu^2/M_j^2$. For
$r_{\mu}^j\ll$ 1 we have
\beq
    f_{H}(r) \simeq- \ln r - 7/6,~~
    f_A (r) \simeq \ln r +11/6, ~~
    f_{H^\pm} (r) \simeq -1/6.
    \label{oneloopintegralsapprox3}
\eeq
For the main two-loop contributions, we have 
\beq
    \Delta a_\mu^{\mbox{$\scriptscriptstyle{\rm 2HDM}$}}({\rm 2-loop})
    = \frac{G_F \, m_{\mu}^2}{4 \pi^2 \sqrt{2}} \, \frac{\alpha_{\rm em}}{\pi}
    \, \sum_{i,f}  N^c_f  \, Q_f^2  \,  y_{\mu}^i  \, y_{f}^i \,  r_{f}^i \,  g_i(r_{f}^i),
\label{barr-zee}
\end{equation}
where $i = H,~ A$, and $m_f$, $Q_f$ and $N^c_f$ are the mass,
electric charge and the number of color degrees of freedom of the
fermion $f$ in the loop. The functions $g_i(r)$ are gievn by \cite{mu2h2-1,mu2h2-2,mu2h2-3}
\begin{eqnarray}
    && g_{h,H}(r) = \int_0^1 \! dx \, \frac{2x (1-x)-1}{x(1-x)-r} \ln
    \frac{x(1-x)}{r}, \\
    && g_{A}(r) = \int_0^1 \! dx \, \frac{1}{x(1-x)-r} \ln
    \frac{x(1-x)}{r}.
\end{eqnarray}
The contributions of $H$ and $A$ to $\Delta a_\mu$
are positive (negative) at one-loop level and negative (positive) at the two-loop level. 
Since $m^2_f/m^2_\mu$ easily
overcomes the loop suppression factor $\alpha/\pi$, the two-loop
contributions can be larger than one-loop ones. 
As a result, the L2HDM can enhance the value of $\Delta a_\mu$ for $m_A < m_H$.
In \cite{2110.13238} the authors presented an extension of the GM2Calc software to calculate the muon $g-2$ of 2HDM precisely.

Because of the large lepton Yukawa couplings, the L2HDM can give sizable corrections to the $Z$ and $\tau$ decays, and thus be
constrained by the measured values of LFU of the $Z$-boson \cite{zexp} 
\begin{eqnarray} \label{lu-zdecay}
{\Gamma_{Z\to \mu^+ \mu^-}\over \Gamma_{Z\to e^+ e^- }} &=& 1.0009 \pm 0.0028
\,,\\ 
{\Gamma_{Z\to \tau^+ \tau^- }\over \Gamma_{Z\to e^+ e^- }} &=& 1.0019 \pm 0.0032
\,,
\end{eqnarray}
and $\tau$ decays \cite{tauexp},
\begin{eqnarray} \label{hfag-data}
&&
\left( g_\tau \over g_\mu \right) =1.0011 \pm 0.0015,~~
\left( g_\tau \over g_e \right) = 1.0029 \pm 0.0015,~~ 
\left( g_\mu \over g_e \right) = 1.0018 \pm 0.0014, 
\nonumber\\
&&
\left( g_\tau \over g_\mu \right)_\pi = 0.9963 \pm 0.0027, \quad
\left( g_\tau \over g_\mu \right)_K = 0.9858 \pm 0.0071.
\end{eqnarray}
Here the first three ratios are defined as 
\begin{eqnarray} 
&&
\left( g_\tau \over g_\mu \right)^2 \equiv \bar{\Gamma}(\tau\to e
\nu\bar{\nu})/\bar{\Gamma}(\mu\to e \nu\bar{\nu}),\\
&&
\left( g_\tau \over g_e \right)^2  \equiv \bar{\Gamma}(\tau\to \mu
\nu\bar{\nu})/\bar{\Gamma}(\mu\to e \nu\bar{\nu}), \\
&&
\left( g_\mu \over g_e \right)^2  \equiv \bar{\Gamma}(\tau\to \mu
\nu\bar{\nu})/\bar{\Gamma}(\tau\to e \nu\bar{\nu}).
\end{eqnarray} 
and the last two ratios are from semi-hadronic processes $\tau \to \pi/K \nu$ and $\pi/K \to \mu \nu$.
$\bar{\Gamma}$ denotes the partial width
normalized to its SM value. The correlation matrix for the above five observables is
\begin{equation} \label{hfag-corr}
\left(
\begin{array}{ccccc}
1 & +0.53 & -0.49 & +0.24 & +0.12 \\
+0.53  & 1     &  + 0.48 & +0.26    & +0.10 \\
-0.49  & +0.48  & 1       &   +0.02 & -0.02 \\
+0.24  & +0.26  & +0.02  &     1    &     +0.05 \\
+0.12  & +0.10  & -0.02  &  +0.05  &   1 
\end{array} \right) .
\end{equation}
The theoretical values of the ratios in the L2HDM are given as 
\begin{eqnarray} \label{deltas-data}
&&
\left( g_\tau \over g_\mu \right) \approx 1+ \delta_{\rm loop}, \quad
\left( g_\tau \over g_e \right) \approx 1+ \delta_{\rm tree}+ \delta_{\rm loop}, \quad
\left( g_\mu \over g_e \right) \approx 1+ \delta_{\rm tree}, 
\nonumber\\
&&
\left( g_\tau \over g_\mu \right)_\pi \approx 1+ \delta_{\rm loop}, \quad
\left( g_\tau \over g_\mu \right)_K \approx 1+ \delta_{\rm loop} .
\end{eqnarray}
Here $\delta_{\rm tree}$ and $\delta_{\rm loop}$ are respectively corrections from
the tree-level diagrams mediated by $H^\pm$ and the one-loop diagrams involved $H$, $A$ and $H^\pm$ \cite{mu2h10,mu2h16},
\begin{eqnarray} 
\delta_{\rm tree} &=& {m_\tau^2 m_\mu^2 \over 8 m^4_{H^\pm}} t^4_\beta
- {m_\mu^2 \over m^2_{H^\pm}} t^2_\beta {g(m_\mu^2/m^2_\tau) \over f(m_\mu^2/m_\tau^2)}, \label{treetau}\\
\delta_{\rm loop} &=& {1 \over 16 \pi^2} { m_\tau^2 \over v^2}  t^2_\beta
\left[1 + {1\over4} \left( H(x_A) + s^2_{\beta-\alpha} H(x_H) + c^2_{\beta-\alpha} H(x_h)\right)
\right]\,, 
\end{eqnarray}
where $f(x)\equiv 1-8x+8x^3-x^4-12x^2 \ln(x)$, $g(x)\equiv 1+9x-9x^2-x^3+6x(1+x)\ln(x)$ and
$H(x_\phi) \equiv \ln(x_\phi) (1+x_\phi)/(1-x_\phi)$ with $x_\phi=m_\phi^2/m_{H^{\pm}}^2$.

\begin{figure}[tb]
\begin{center}
  \epsfig{file=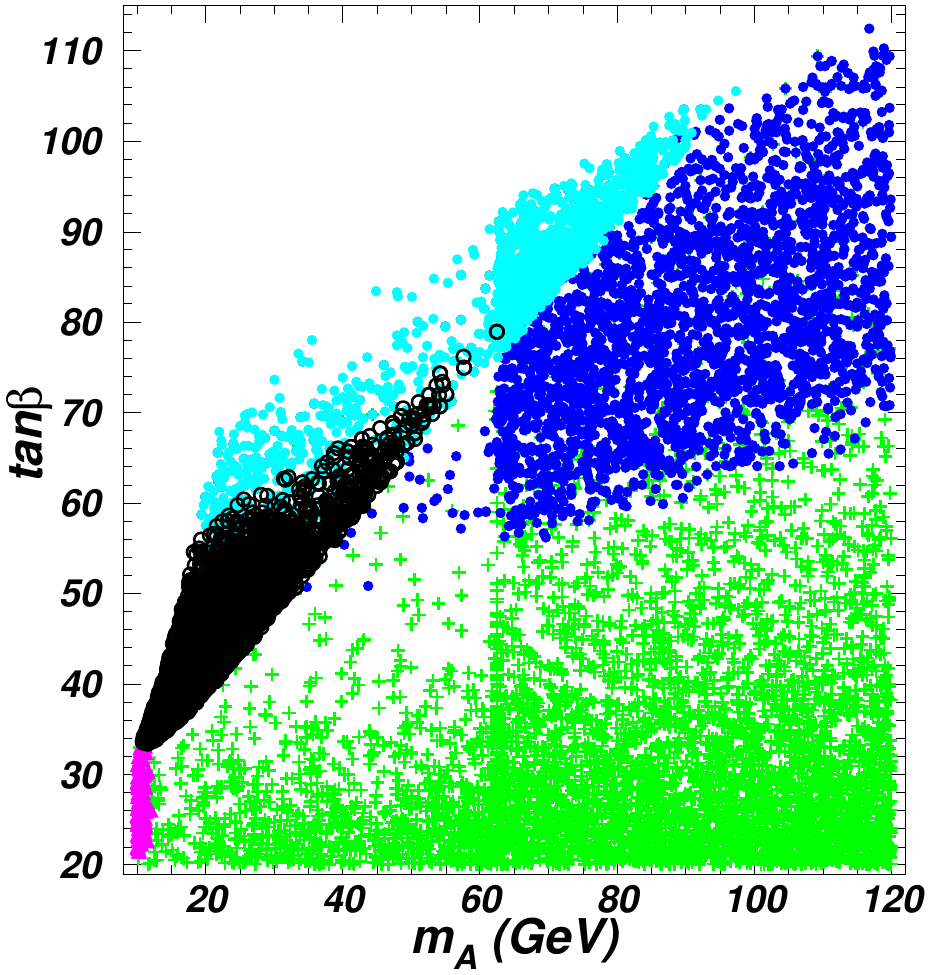,height=5.7cm}
  \epsfig{file=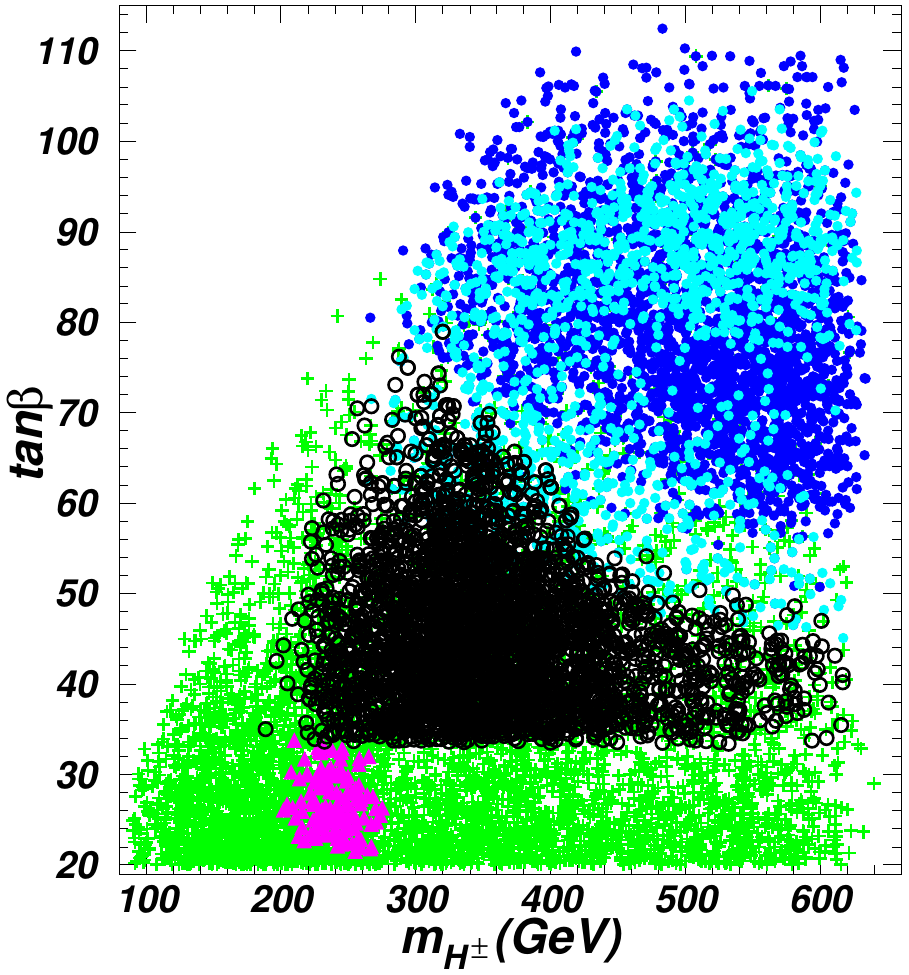,height=5.7cm}
  \epsfig{file=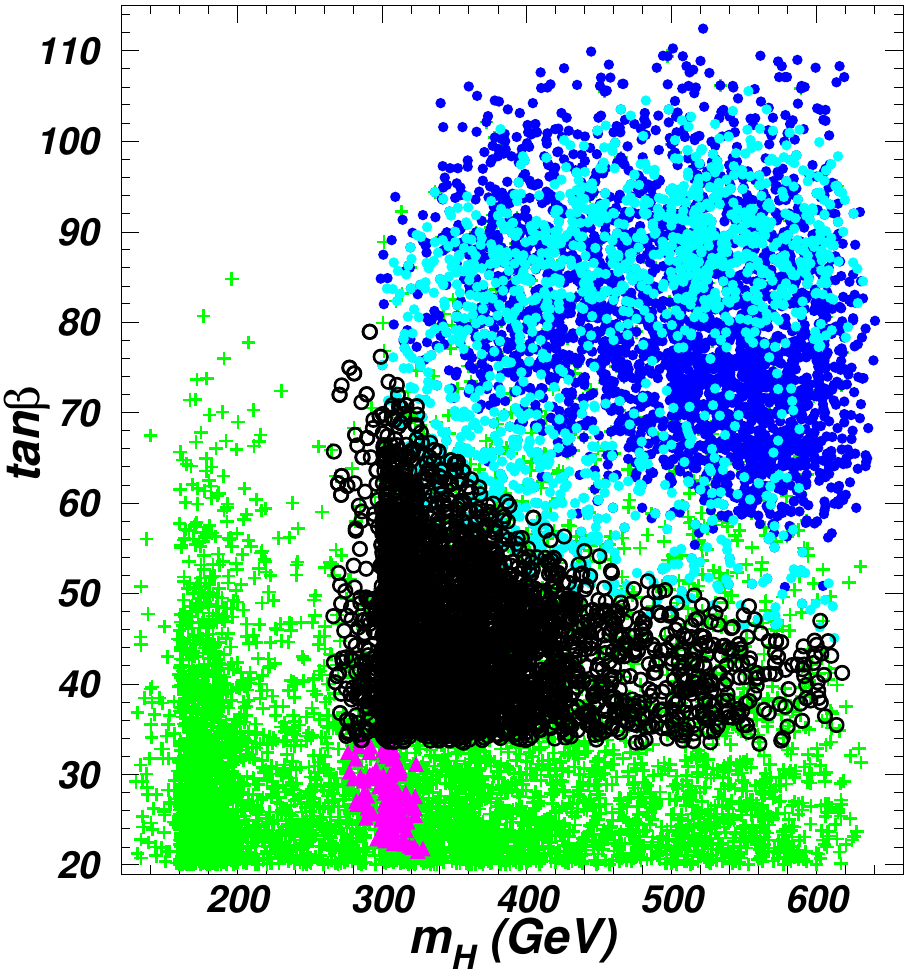,height=5.7cm}
  \epsfig{file=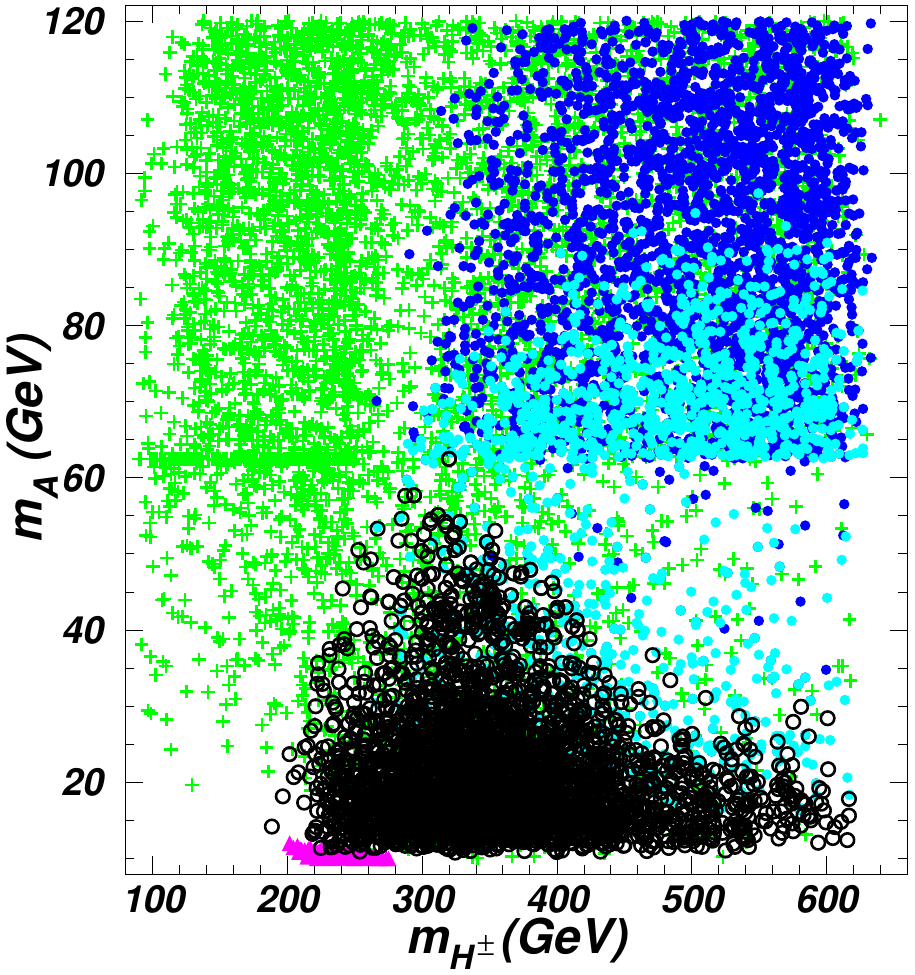,height=5.7cm}
  \epsfig{file=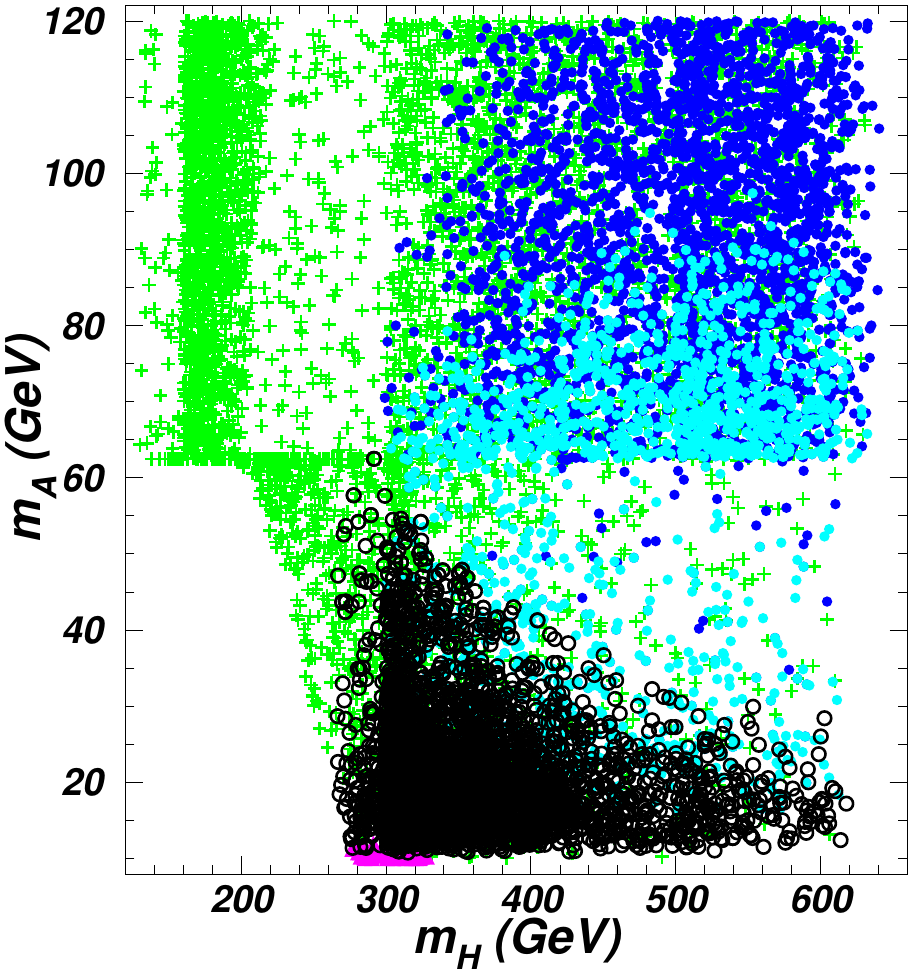,height=5.7cm}
  \epsfig{file=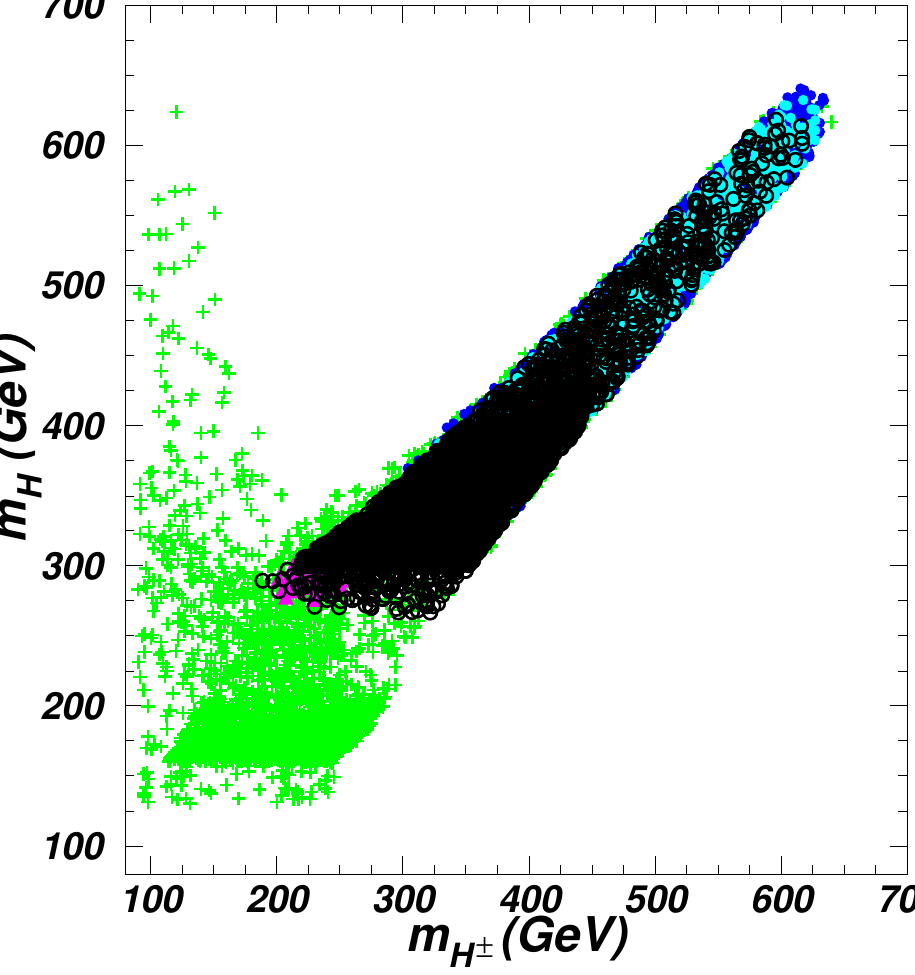,height=5.7cm}
\end{center}
\vspace{-0.25cm} \caption{All the samples are allowed by the constraints of "pre-muon $g-2$", taken from \cite{mu2h17}.
The triangles (pink) are excluded by the Br$(B_s\to \mu^+\mu^-)$ data,  
the light bullets (sky blue) and dark bullets (royal blue) are excluded by LFU in $Z$ decay.
The light (dark) bullets can (cannot) explain the muon $g-2$ anomaly. 
The circles (black) are allowed by the constraints from the muon $g-2$, "pre-muon $g-2$", 
the LFU in $Z$ decay, and Br$(B_s\to \mu^+\mu^-)$.}
\label{mug2}
\end{figure}

The experimental value of $\left( g_\tau \over g_e \right)$ has an approximately $2\sigma$ positive 
deviation from the SM. In  
the L2HDM, the tree-level diagram mediated by $H^{\pm}$ gives negative contribution to the
decay $\tau\to \mu\nu\bar{\nu}$, as shown in Eq. (\ref{treetau}), which tends to raise the discrepancy
in the LFU in  $\tau$ decays.
In \cite{mu2h17}, a global fit to the LFU data from $\tau$ decays and the 125 GeV Higgs signal data was performed, 
requiring $\chi^2-\chi^2_{\rm min} \leq 6.18$ with $\chi^2_{\rm min}$ denoting the minimum of $\chi^2$. 
Fig. \ref{mug2} shows the surviving samples satisfying the constraints of "pre-muon $g-2$" (denoting the theory, 
the oblique parameters, the exclusion limits from the searches for Higgs at LEP,
the signal data of the 125 GeV Higgs, LFU in $\tau$ decays,
and the exclusion limits from $h\to AA$ channels at LHC). The LFU in $Z$ decays can exclude most of samples in the region of large $m_{H^\pm}$ ($m_{H}$) and $\tan\beta$. 
In addition to a large $\tan\beta$, one-loop diagrams can give a sizable correction to the 
LFU in $Z$ decays for $m_A<m_{H^\pm}~(m_H)$.
The oblique parameters favor $H$ and $H^\pm$ to have a small splitting mass for a light $A$.  
Many samples in the regions of $m_A$ around 10 GeV and $m_{H^\pm}<$ 300 GeV are excluded by 
the measurement of Br$(B_s\to \mu^+\mu^-)$. This is because that the decay $B_s\to \mu^+\mu^-$ can
get sizable corrections from the $A$-exchange diagrams for a very small $m_A$.
 
Under various theoretical and experimental constraints, the L2HDM can explain
the muon $g-2$ anomaly in the regions of 32 $<\tan\beta<$ 80, 
10 GeV $<m_A<$ 65 GeV,
260 GeV $<m_H<$ 620 GeV, and 180 GeV $<m_{H^\pm}<$ 620 GeV.  
 Because the contributions of $A$ and $H$ to the muon $g-2$ anomaly are respectively positive and negative, 
the mass splitting between $A$ and $H$ is required to be large to explain the muon $g-2$ anomaly,
as shown in the lower-middle panel of Fig. \ref{mug2}.

\begin{figure}[thb]
\begin{center}
  \epsfig{file=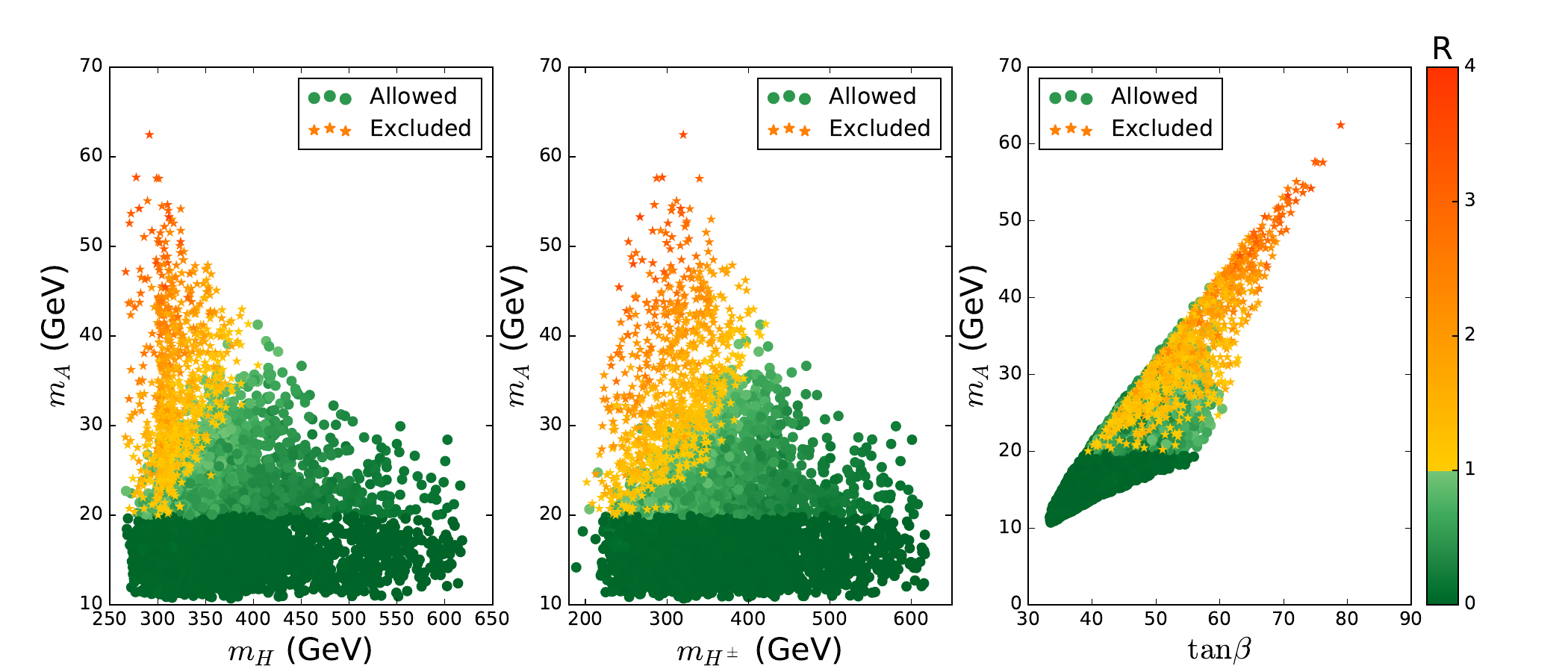,width=15cm}
\end{center} 
\vspace{-0.5cm} 
\caption{All the samples are allowed by the constraints from the muon $g-2$, "pre-muon $g-2$", the LFU 
in $Z$ decay, and Br$(B_s\to \mu^+\mu^-)$, taken from \cite{mu2h17}. 
The orange stars and green dots are excluded and allowed by the LHC Run-2 
data at 95\% confidence level, respectively. }
\label{lhc2}
\end{figure}

 The muon $g-2$ explanation requires a
large $\tan\beta$ which will sizably suppress the quark Yukawa couplings of $H$, $A$ and $H^\pm$.
Therefore, these extra Higgs bosons are dominantly produced at the LHC via
the following electroweak processes:
\begin{align}
pp\to & W^{\pm *} \to H^\pm A, \label{process1}\\
pp\to &       Z^*/\gamma^* \to HA, \label{process2}\\
pp\to & W^{\pm *} \to H^\pm H, \label{process3}\\
pp\to & Z^*/\gamma^* \to H^+H^-. \label{process4}
\end{align}
The main decay modes of the Higgs bosons are 
\begin{align}
A\to & \tau^+\tau^-,~\mu^+\mu^-,\cdots\cdots,\\
H\to & \tau^+\tau^-,~ZA,\cdots\cdots,\\
H^\pm\to & \tau^\pm\nu,~W^\pm A,\cdots\cdots.
\end{align}
Therefore, in the parameter space favored by the muon $g-2$ explanation, the L2HDM will mainly 
produce multi-lepton signature at the LHC, especially the multi-$\tau$ signature.

The study in \cite{mu2h17} used all the analysis for the 13 TeV LHC in 
\texttt{CheckMATE 2.0.7}~\cite{Dercks:2016npn} and the multi-lepton searches for 
electroweakino~\cite{Sirunyan:2017zss,Sirunyan:2017qaj,Sirunyan:2017lae,Sirunyan:2018ubx,Aaboud:2017nhr}
 to constrain the parameter space.
The surviving samples are shown in Fig. \ref{lhc2} in which \texttt{R} $>1$ denotes
 that the corresponding samples are excluded at 95\% confidence level.
The searches for multi-leptons at the 13 TeV LHC shrink $m_A$ from [10, 65] GeV 
to [10, 44] GeV and $\tan\beta$ from [32, 80] to [32, 60]. 
The main constraint is given by the search for electroweak production of charginos and neutralinos 
in multi-lepton final states \cite{Sirunyan:2017lae}. 

For relatively large $m_{H}$ and $m_{H^\pm}$, the production cross sections of extra Higgs bosons 
are small enough to escape the limits of direct searches at the LHC.
For a light $A$, the $\tau$ leptons from the decays of $A$ produced in Eq. (\ref{process1}) 
and Eq. (\ref{process2}) are too soft to be distinguished at detectors, and
 the $\tau$ leptons from $A$ produced in $H/H^\pm$ decays are collinear because of the large 
mass splitting between $A$ and $H/H^\pm$. Thus, in the very low $m_A$ region, 
the acceptance of above signal region quickly decreases and the limits of direct searches 
can be easily satisfied. 

\subsection{Solution of muon $g-2$ and $\tau$ decays}
The L2HDM can give a simple explanation for the muon $g-2$, but raise the discrepancy
in the LFU in  $\tau$ decays. Therefore, to explain the 
muon $g-2$ and LFU of $\tau$ decays simultaneously, other models need to be considered. 

\subsubsection{Lepton specific inert 2HDM}
In this model the $Z_2$ symmetry-breaking lepton Yukawa interactions of $\Phi_2$ are added 
to the inert 2HDM \cite{mueg,2104.03227}
\bea 
\label{phi2coupling} - {\cal L} &=& \frac{\sqrt{2}m_e}{v}\, \kappa_e \,\overline{L}_{1L} \, {\Phi}_2
\, e_R  \, +\frac{\sqrt{2}m_\mu}{v}\, \kappa_\mu\, \overline{L}_{2L} \, {\Phi}_2
\,\mu_R \, \nonumber\\
&&+\frac{\sqrt{2}m_\tau}{v}\, \kappa_\tau \,\overline{L}_{3L} \, {\Phi}_2
\,\tau_R\, + \, \mbox{h.c.}\,. 
\eea
In this way the extra Higgs bosons ($H$, $A$, and $H^\pm$) acquire couplings 
to the leptons while have no couplings to the quarks.

In this model, $\left( g_\tau \over g_e \right)^2$ is given by 
\beq \label{deltas-data}
\left( g_\tau \over g_e \right)^2   \approx \frac{1+ 2\delta_{\rm tree}+ 2\delta^\tau_{\rm loop}}{1+2\delta^\mu_{\rm loop}}.
\eeq 
Here $\delta_{\rm tree}$ and $\delta_{\rm loop}^{\tau,\mu}$ are respectively corrections from
the tree-level diagrams mediated by $H^\pm$ and the one-loop diagrams involving 
$H$, $A$ and $H^\pm$, given by \cite{mu2h10,mu2h16,mu2h17}
\begin{eqnarray} \label{tree-tau}
\delta_{\rm tree} &=& {m_\tau^2 m_\mu^2 \over 8 m^4_{H^\pm}} \kappa^2_\tau \kappa^2_\mu
- {m_\mu^2 \over m^2_{H^\pm}} \kappa_\tau \kappa_\mu {g(m_\mu^2/m^2_\tau) \over f(m_\mu^2/m_\tau^2)}, \\
\delta_{\rm loop}^{\tau,\mu} &=& {1 \over 16 \pi^2} { m_{\tau,\mu}^2 \over v^2}  \kappa^2_{\tau,\mu}
\left[1 + {1\over4} \left( H(x_A)  + H(x_H)\right)
\right]\,. 
\end{eqnarray}

The model gives the one-loop contributions to muon $g-2$ \cite{mu2h1-1,mu2h1-2,mu2h1-3}
 \beq
    \Delta a_\mu^{\mbox{$\scriptscriptstyle{\rm 2HDM}$}}({\rm 1loop}) =
    \frac{m_\mu^2}{8 \pi^2 v^2} \, \sum_i
     \kappa_\mu^2 \, r_{\mu}^i \, F_j(r_{\mu}^i),
\label{amuoneloop}
\end{equation}
where $i = H,~ A ,~ H^\pm$ and $r_{\mu}^ i =  m_\mu^2/M_j^2$. 

The contributions of the two-loop diagrams are 
\beq
    \Delta a_\mu^{\mbox{$\scriptscriptstyle{\rm 2HDM}$}}({\rm 2loop})
    = \frac{m_{\mu}^2}{8 \pi^2 v^2} \, \frac{\alpha_{\rm em}}{\pi}
    \, \sum_{i,\ell} \, Q_\ell^2  \,  \kappa_\mu  \, \kappa_\ell \,  r_{\ell}^i \,  G_i(r_{\ell}^i),
\label{barr-zee}
\end{equation}
where $i = H,~ A$, $\ell=\tau$, and $m_\ell$ and $Q_\ell$ are the mass and
electric charge of the lepton $\ell$ in the loop.

This model was also used to discuss the electron $g-2$ anomaly, and the calculations are similar to 
the muon $g-2$.
The value from the measurement of the fine-structure constant using $^{133}$Cs atoms at 
Berkeley \cite{alpha-exp} makes the electron $g-2$ to have
$2.4\sigma$ deviation from the SM prediction \cite{eg2-2.4-1,eg2-2.4-2}, 
\bea
\Delta a_e=a_e^{\rm exp}-a_e^{\rm SM}=(-87\pm36)\times10^{-14}.
\eea
However, the newest experimental result of the fine-structure constant using $^{87}$Rb atoms at Laboratoire
Kastler Brossel gives a value of $a_e$  which agrees well with the the SM value \cite{eg2-lkb}.
So far, the discrepancy between these two experimental results is not clear. If
the Berkeley $^{133}$Cs experiment result turns out to be the real story, it will be challenging 
to explain muon and electron $g-2$ simultaneously
since the two effects have opposite sign. In  \cite{mueg}, the Berkeley $^{133}$Cs experiment data was 
used and this model was found to give explanation for the muon and electron $g-2$ (for the 
explanation in other popular models like the minimal supersymmetry, see, e.g., \cite{Li:2021koa}).   

\begin{figure*}[tb]
\begin{center}
\includegraphics[width=8cm]{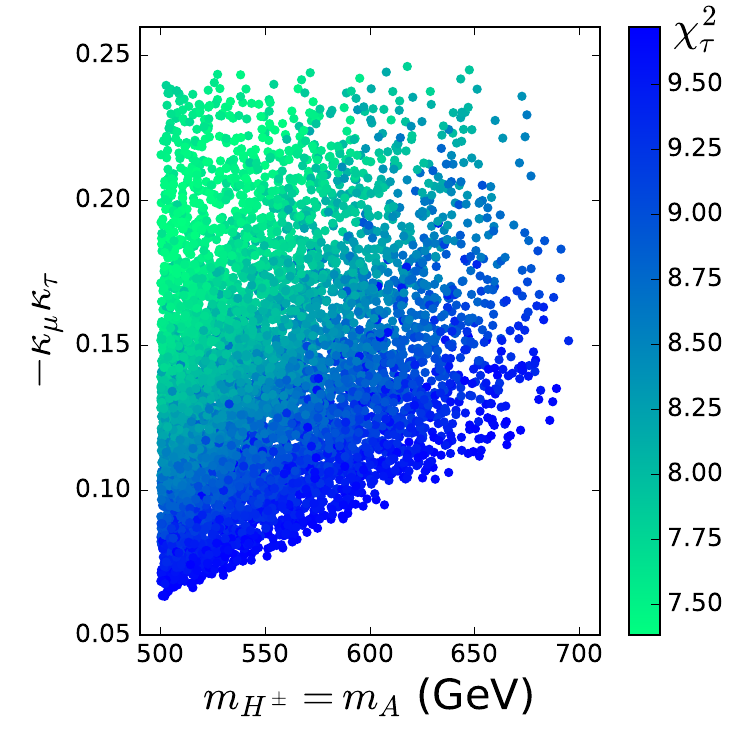}
\end{center}
\vspace{-1.0cm}\caption{In the lepton-specific inert 2HDM, the surviving samples fit the data of LFU in $\tau$ decay within the $2\sigma$ range, taken from \cite{mueg}.}
\label{tauyes}
\end{figure*}

\begin{figure}[tb]
\begin{center}
\includegraphics[width=16cm]{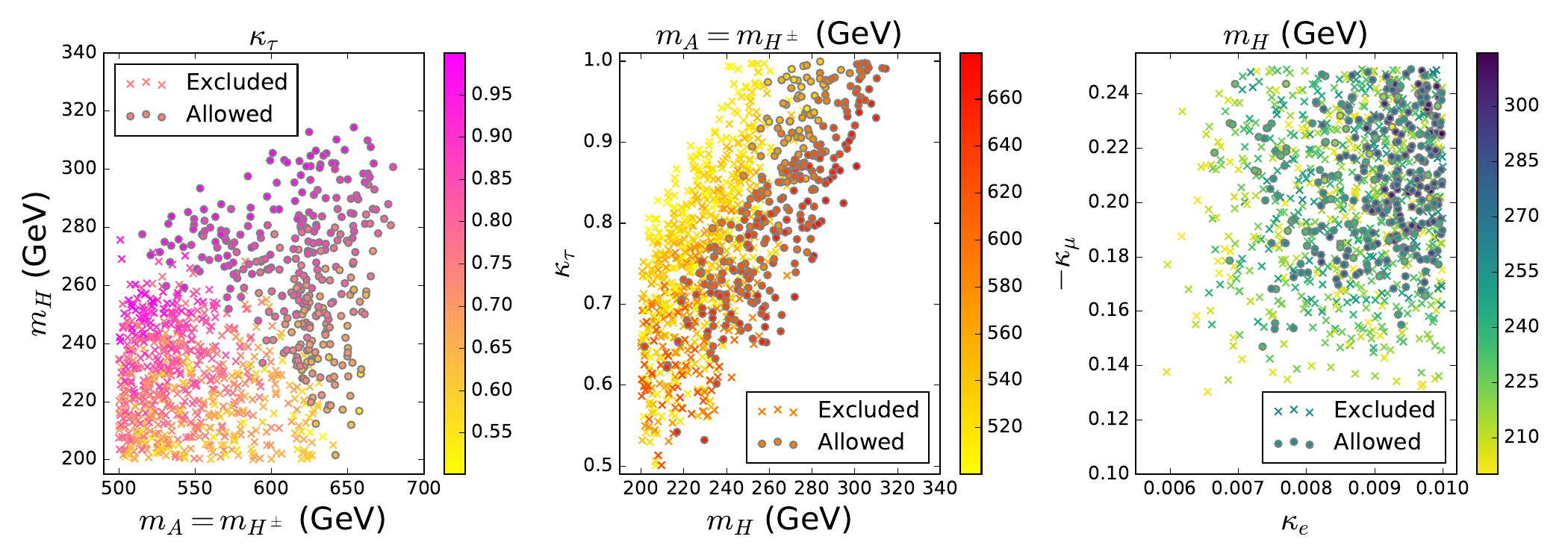}
\end{center}
\vspace*{-0.7cm}
\caption{All the samples are allowed by the constraints of theory, the oblique parameters, $\Delta a_\mu$, $\Delta a_e$, 
the data of LFU in $\tau$ decays, and $Z$ decay, taken from \cite{mueg}. The bullets and crosses are respectively allowed and excluded by
 the direct search limits from the LHC at 95\% confidence level. The colors denote $\kappa_{\tau}$, $m_A$ and $m_{H}$ in left, middle, and right panel, respectively.
 }
\label{muelhc}
\end{figure}

Taking 
\beq
\kappa_\mu < 0,~~~ \kappa_\tau > 0,~~~ \kappa_e > 0,
\eeq
then $\delta_{\rm tree}$ has a positive value because of the opposite signs of $\kappa_\mu$ and $\kappa_\tau$.
Thus, the model can enhance $\left( g_\tau \over g_e \right)$ and give a better fit to the data of 
the LFU in the $\tau$ decays.
The contributions of $H$ ($A$) to the muon $g-2$
are positive (negative) at the two-loop level and positive (negative) at one-loop level. 
For the electron $g-2$, the contributions of $H$ ($A$) 
are negative (positive) at the two-loop level and positive (negative) at one-loop level.
Fig. \ref{tauyes} shows the surviving samples with $\chi_\tau^2<$ 9.72, which means to fit the data 
of LFU in $\tau$ decays within $2\sigma$ range. 
 Fig. \ref{tauyes} shows that $\chi_\tau^2$ can be as low as 7.4, which is much smaller than 
the SM prediction (12.25). 
Fig. \ref{muelhc} shows that after imposing the constraints of theory, the oblique parameters, 
the $Z$ decay, and the direct searches at LHC,
the model can simultaneously explain the anomalies of $\Delta a_\mu$, $\Delta a_e$ and LFU in 
the $\tau$ decay within $2\sigma$ range
 in a large parameter space of 200 GeV $<m_H<$ 320 GeV, 
500 GeV $<m_A=m_{H^{\pm}}<680$ GeV, 0.0066 $<\kappa_e<0.01$,  
-0.25 $<\kappa_\mu<-0.147$, and 0.53 $<\kappa_\tau<1.0$.

\subsubsection{$\mu$-$\tau$-philic Higgs doublet model}
In this model an exact discrete $Z_4$ symmetry is imposed, and the $Z_4$ charge assignment 
is shown in Table~\ref{tab:matter} \cite{190410908}.
The scalar potential is given as
\begin{eqnarray}  \mathrm{V}   &=&   Y_1
(\Phi_1^{\dagger} \Phi_1) + Y_2 (\Phi_2^{\dagger}
\Phi_2)+ \frac{\lambda_1}{2}  (\Phi_1^{\dagger}  \Phi_1)^2 +
\frac{\lambda_2}{2}  (\Phi_2^{\dagger} \Phi_2)^2  \nonumber \\
&&+ \lambda_3
(\Phi_1^{\dagger}  \Phi_1)(\Phi_2^{\dagger} \Phi_2) + \lambda_4
(\Phi_1^{\dagger} 
\Phi_2)(\Phi_2^{  \dagger} \Phi_1)   + \left[\frac{\lambda_5}{2}   (\Phi_1^{\dagger} \Phi_2)^2 + \rm
h.c.\right].
\end{eqnarray}
The vev of the $\Phi_1$ field is $v$=246 GeV, while the $\Phi_2$ field has zero VeV. 
The fermions obtain masses via the Yukawa interactions with $\Phi_1$
 \beq \label{yukawacoupling} - {\cal L} = y_u\overline{Q}_L \,
\tilde{{ \phi}}_1 \,U_R + y_d\overline{Q}_L\,{\phi}_1 \, D_R +  y_\ell\overline{L}_L \, {\phi}_1
\, E_R + \mbox{h.c.}. \eeq

\begin{table}
\caption{The $Z_4$ charge assignment of $\mu$-$\tau$-philic 2HDM.}
\label{tab:matter}
\begin{center}
\begin{tabular}{cccccccccccc}
\hline
&~$\Phi_1$~&~$\Phi_2$~&~$Q_L^{i}$~&~$U_R^i$~&~$D_R^i$~&~$L_L^e$~&~$L_L^\mu$~&~$L_L^\tau$~&~$e_R$~&~$\mu_R$~&~$\tau_R$~\\ \hline
    ~~$Z_4$~& $1$  & -1 &     1    & 1   &  1    &  1      & $\textrm{i}$      & ${\rm -i}$     & 1   & $\textrm{i}$    & $\textrm{-i}$      \\ \hline
\end{tabular}
\end{center}
\end{table}

The $Z_4$ symmetry allows $\Phi_2$ to have $\mu$-$\tau$ interactions \cite{190410908}
\bea\label{lepyukawa2}
- {\cal L}_{\rm LFV} & = &  \sqrt{2}~\rho_{\mu\tau}  \,\overline{L^\mu_{L}} \,  {\phi}_2
\,\tau_R  \, +  \sqrt{2}~\rho_{\tau\mu}\, \overline{L^\tau_{L}} \, {\phi}_2
\,\mu_R \, + \,  \mbox{h.c.}\,. \eea
From these interactions we can obtain the $\mu$-$\tau$ lepton flavor violation (LFV) couplings 
of $H$, $A$, and $H^\pm$.

The model gives new contribution to $\Delta a_{\mu}$ via the one-loop diagrams containing 
the $\mu$-$\tau$ LFV coupling of $H$ and $A$ \cite{10010434,taumug2-2,taumug2-3} 
\bea
  \Delta a_{\mu} = \frac{m_\mu m_\tau  \rho^2}{8\pi^2}
  \left[\frac{ (\log\frac{m_H^2}{m_\tau^2} - \frac{3}{2})}{m_H^2}
  -\frac{\log( \frac{m_A^2}{m_\tau^2}-\frac{3}{2})}{m_A^2}
\right],
  \label{mua1}
\eea
which shows that the contributions of $H$ and $A$ are respectively positive and negative.
\begin{figure}[tb]
\begin{center}
 \epsfig{file=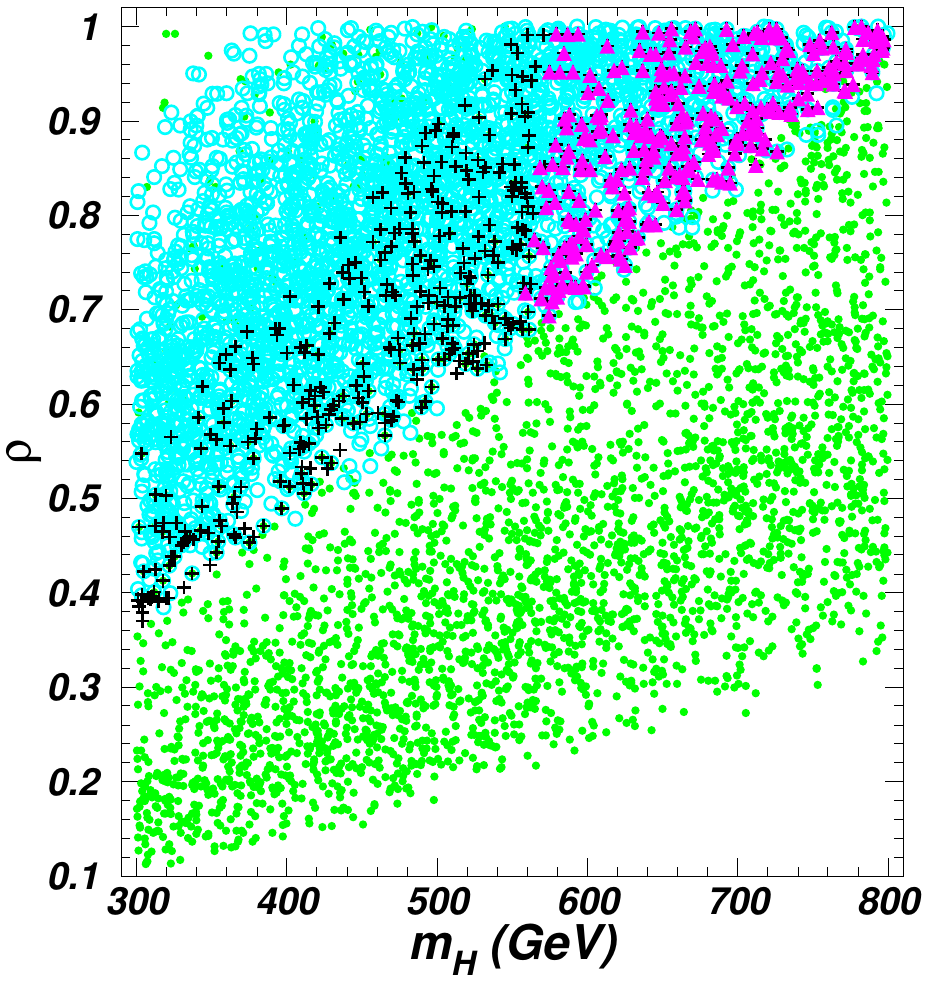,height=6.0cm}
\epsfig{file=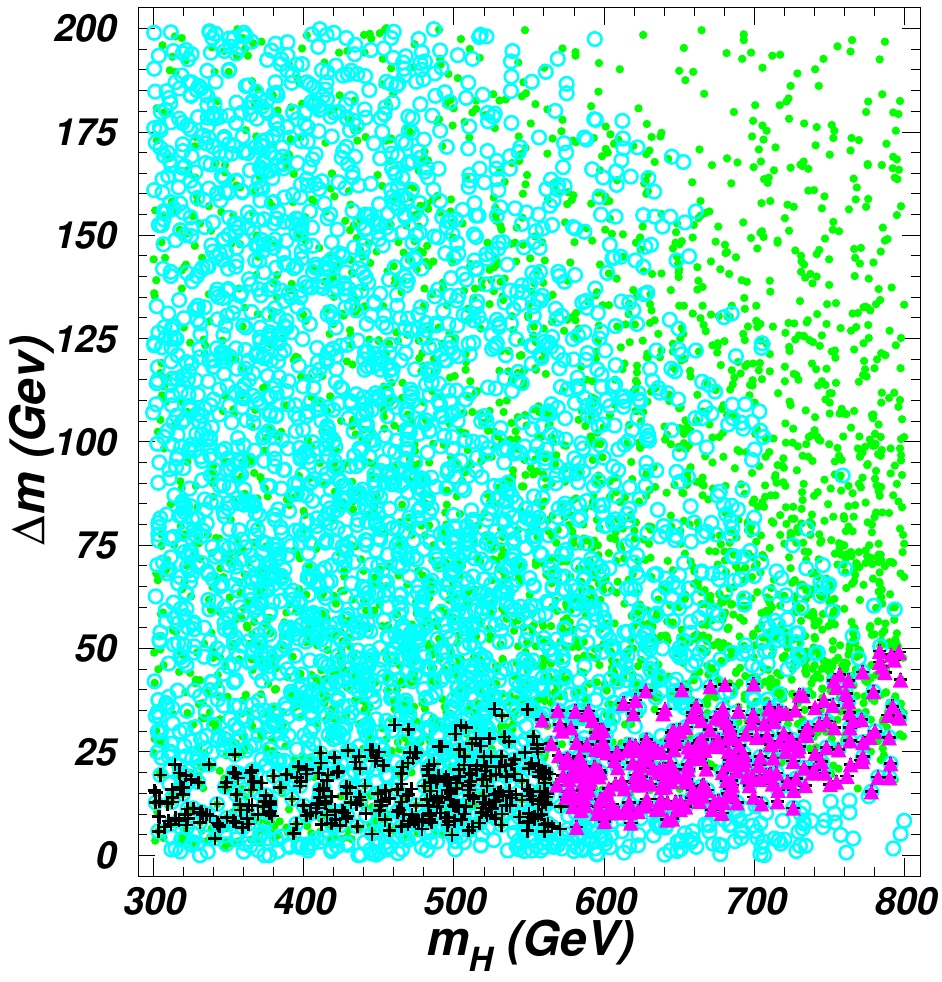,height=6.0cm}
\end{center}
\vspace{-0.5cm} 
\caption{The surviving samples of the $\mu$-$\tau$-philic 2HDM allowed 
by the constraints of the theory,
 the oblique parameters, and the $Z$ decays, taken from \cite{1908.03755}. 
The bullets (green) samples are within the $2\sigma$ range of muon $g-2$ and
 the circles (blue) are within the $2\sigma$ range of LFU in the $\tau$ decays. 
The triangles (purple) and  pluses (black)
are within the $2\sigma$ ranges of both muon $g-2$ and LFU in the $\tau$ decays, 
and the former are allowed by the limits of the direct searches at the LHC, 
while the latter are excluded. Here $\Delta m\equiv m_A -m_H$ with $m_A=m_{H^\pm}$.}
\label{mg2tau}
\end{figure}

In this model, $\left( g_\tau \over g_e \right)^2$ is given as \cite{190410908,1908.03755}
\begin{eqnarray} 
&&\left( g_\tau \over g_e \right)^2  =  (1+\delta_{\rm loop}^\tau)^2+\frac{\delta_{\rm tree}}{(1+\delta_{\rm loop}^\mu)^2},
\end{eqnarray}
where the flavor of final neutrino and anti-neutrino states is summed up, and $\delta_{\rm tree}$ 
is from the tree-level diagram mediated by the charged Higgs
\beq
\delta_{\rm tree}=4\frac{m_W^4\rho^4}{g^4 m_{H^{\pm}}^4},
\label{delta-tree}
\eeq
with $\rho_{\mu\tau}=\rho_{\tau\mu}=\rho$,
and $\delta_{\rm loop}^\mu$ and $\delta_{\rm loop}^\tau$ are the corrections to vertices 
$W\bar{\nu_{\mu}}\mu$ and $W\bar{\nu_{\tau}}\tau$ from the one-loop diagrams
 involving $A$, $H$, and $H^\pm$
\beq
\delta_{\rm loop}^\tau=\delta_{\rm loop}^\mu={1 \over 16 \pi^2} {\rho^2} 
\left[1 + {1\over4} \left( H(x_A) +  H(x_H) \right)
\right]\,. 
\label{delta-loop}
\eeq
Since $\delta_{\rm tree}$ is positive, the model can enhance $\left( g_\tau \over g_e \right)$ and give a better fit to the data of the LFU in the $\tau$ decays.

Fig. \ref{mg2tau} shows that after considering the constraints from theory, 
the oblique parameters and the $Z$ decay,
the model can simultaneously explain $\Delta a_\mu$ and LFU in the $\tau$ decays 
in the parameter space with 300 GeV $<m_H<$ 800 GeV and $\Delta m<$ 50 GeV. 
For such small mass splitting between $m_A$ ($m_{H^\pm}$) and $m_H$,
$H$, $A$ and $H^\pm$ will mainly decay into $\tau\mu$, $\tau\nu_\mu$, and $\mu\nu_\tau$. 
The limits of direct searches at the LHC exclude the region $m_H< 560$ GeV and require $\rho>$ 0.68.
Also Refs. \cite{1907.09845,2104.03242} discussed the implications of the muon $g-2$ anomaly on the model.

In the discussions above, $m_A$ and $m_{H^\pm}$ are chosen to have degenerate mass, which is disfavored by the CDF $W$-mass. The study in Ref. \cite{mw2h-4} found that
combined with relevant theoretical and experimental constraints, the mass splittings among 
$H$, $A$ and $H^\pm$ of the model are stringently constrained in the region
 simultaneously explaining the $W$-mass, muon $g-2$ and LFU in $\tau$ decays, i.e., 
10 GeV $<m_A-m_H<$ 75 GeV, 65 GeV $<m_{H^\pm}-m_A<$ 100 GeV, 85 GeV $<m_{H^\pm}-m_H<$ 125 GeV
( -150 GeV $<m_{H^\pm}-m_A<$ -85 GeV, -105 GeV $<m_{H^\pm}-m_H<$ -55 GeV).

\subsection{Other 2HDMs and muon $g-2$}
In \cite{1705.01469} the authors proposed a muon specific 2HDM
in which  extra Higgs boson couplings to muon are enhanced by a factor of $\tan\beta$, while
their couplings to the other fermions are suppressed by $\cot\beta$. Thus, the
model can explain the muon $g-2$ anomaly by the contributions of the one-loop diagram for a very large $\tan\beta$, and
weaken the constraints of the $\tau$ decays because of the cancellation of $\tan\beta$ between the tau and the muon Yukawa couplings to the charged Higgs.
The study in \cite{mu2h16-2} considered a perturbed lepton-specific 2HDM in which the sign of extra Higgs boson couplings to tau are flipped.
Similar to the lepton-specific inert 2HDM, the model can accomodate the muon $g-2$ anomaly and the $\tau$ decays, but the electron $g-2$ anomaly is
not simultaneously explained. The muon and electron $g-2$ anomalies can be both explained in general 2HDM. For recent studies see e.g. \cite{2006.01934,2003.03386,2110.01356}.

The L2HDM can be derived from the aligned 2HDM by taking the specific parameter space \cite{aligned2h}.
Therefore, the aligned 2HDM may explain the muon $g-2$ anomaly in a broader parameter space, 
and the large lepton Yukawa couplings of $A$ still play the main role in most of parameter space \cite{a2h0}. 
Some recent studies have been done in \cite{a2h-1,a2h-2,a2h-3,a2h-4,2012.06911}.
In addition to the 2HDM with a $Z_4$ symmetry, the extra Higgs doublet with the 
$\mu$-$\tau$ LFV interactions can be obtained in general 2HDM. 
For recent studies see e.g. \cite{2111.10464,2105.11315,2010.03590,lfv-1,lfv-2,1610.06587,1711.08430}.

\section{Summary} \label{summary} 
From the above review we summarize the following points for the 2HDMs:   
(i) For the popular type-II 2HDM,  
the current direct searches at the LHC excluded a large part of parameter space, 
while still allowing the 125 GeV Higgs to have wrong-sign Yukawa couplings to 
the down-type quarks and leptons. 
If a real singlet scalar DM is added to the type-II 2HDM and the 125 GeV Higgs with wrong-sign 
Yukawa couplings is taken as the portal between the SM sector
and DM, this model can have isospin-violating interactions between DM and nucleons, which can 
relax the constraints from the DM direct detections; 
(ii) As a simpler DM model, in several DM mass ranges the inert 2HDM can produce the 
correct relic density and satisfy the bounds of DM direct detections and direct searches at the LHC; 
(iii) The muon $g-2$ anomaly can be explained in the lepton-specific 2HDM with a light $A$ 
and heavy $H/H^\pm$, but it will raise the discrepancy in the LFU in $\tau$ decays. 
Such a tension may be solved in the 2HDM with some specific muon and tau Yukawa couplings. 
So, compared with low energy supersymmetry (for recent reviews see, e.g., \cite{Wang:2022rfd,Baer:2020kwz}),
the 2HDMs can also do the job of explaining dark matter and muon $g-2$, albeit cannot address the 
naturalness problem. 
     
\addcontentsline{toc}{section}{Acknowledgments}
\acknowledgments
 This work was supported by the National Natural Science Foundation of China 
(NNSFC) under grant Nos. 11975013, 11821505, 12075300 and 12105248, 
by Peng-Huan-Wu Theoretical Physics Innovation Center (12047503),
by the CAS Center for Excellence in Particle Physics (CCEPP), 
and by the Key Research Program of the Chinese Academy of Sciences, Grant NO. XDPB15.

\end{document}